\documentclass{article}

 \usepackage[eandd, final]{neurips_2026}
\usepackage[utf8]{inputenc} % allow utf-8 input
\usepackage[T1]{fontenc}    % use 8-bit T1 fonts
\usepackage{hyperref}       % hyperlinks
\usepackage{url}            % simple URL typesetting
\usepackage{booktabs}       % professional-quality tables
\usepackage{amsfonts}       % blackboard math symbols
\usepackage{nicefrac}       % compact symbols for 1/2, etc.
\usepackage{microtype}      % microtypography
\usepackage{xcolor}         % colors
\usepackage{caption}
\usepackage{graphicx}
\usepackage{amsmath} 
\usepackage{makecell}
\usepackage{multirow}
\usepackage{xcolor}
\usepackage{float}
\usepackage[table]{xcolor}
\usepackage{booktabs}
\usepackage{makecell}
\usepackage{multirow}
\usepackage{soul}
\usepackage{acro}
\usepackage{enumitem}
\usepackage{longtable}
\usepackage{booktabs}
\usepackage{pdflscape}

\DeclareAcronym{rmse}{
  short=RMSE,
  long=Root-Mean-Square error
}
\DeclareAcronym{nrmse}{
  short=NRMSE,
  long=Normalised Root-Mean-Square Error
}

\newcommand{\bench}{\textsc{TokaMark}}
\newcommand{\DatasetSize}{11,573}
\newcommand{\TotalSignals}{39}

\title{\bench: A Comprehensive Benchmark for MAST Tokamak Plasma Models}

\author{%
\textbf{Cécile Rousseau}$^{1}$ \quad \textbf{Samuel Jackson}$^{2}$ \quad \textbf{Rodrigo H. Ordonez-Hurtado}$^{1}$ \\
\textbf{Nicola C. Amorisco}$^{2}$ \quad
\textbf{Tobia Boschi}$^{1}$ \quad \textbf{George K. Holt}$^{3}$ \quad \textbf{Andrea Loreti}$^{2}$ \quad \textbf{Eszter Sz\'ekely}$^{2}$ \\
\textbf{Alexander Whittle}$^{2}$ \quad \textbf{Adriano Agnello}$^{3}$ \quad \textbf{Stanislas Pamela}$^{2}$ \quad \textbf{Alessandra Pascale}$^{1}$ \\
\textbf{Robert Akers}$^{2}$ \quad \textbf{Juan Bernabe Moreno}$^{1}$ \quad \textbf{Sue Thorne}$^{3}$ \quad \textbf{Mykhaylo Zayats}$^{1}$ \\
$^{1}$IBM Research Europe \quad $^{2}$UK Atomic Energy Authority \quad $^{3}$STFC Hartree Centre \\
\texttt{rousseau.cecile@ibm.com, samuel.jackson@ukaea.uk, mykhaylo.zayats1@ibm.com}
}

\begin{document}

\maketitle

\begin{abstract}
  Development and operation of commercially viable fusion energy reactors such as tokamaks require accurate predictions of plasma dynamics from sparse, noisy, and incomplete sensors readings. The complexity of the underlying physics and the heterogeneity of experimental data pose formidable challenges for conventional numerical methods, and highlight the promise of modern data-native approaches. A major obstacle in realizing this potential is, however, the lack of curated, openly available datasets and standardized benchmarks. 
  Existing fusion datasets are scarce, fragmented across institutions, facility-specific, and inconsistently annotated, which limits reproducibility and prevents a fair and scalable comparison of AI approaches. 
  In this paper, we introduce \textbf{\bench}, a structured benchmark to evaluate AI models on real experimental data collected from the Mega Ampere Spherical Tokamak (MAST). \bench\ provides a comprehensive suite of tools designed to unify access to multi-modal fusion data and standardize evaluation protocols. The benchmark includes a curated list of 14 tasks spanning a range of physical mechanisms, exploiting a variety of diagnostics and covering multiple operational use cases. A baseline model is provided to facilitate transparent comparison and validation within a unified framework. By establishing a unified benchmark, \bench\ aims to accelerate progress in data-driven AI-based plasma modeling, contributing to the broader goal of achieving sustainable and stable fusion energy. The dataset, benchmark, documentation, and tooling are open-sourced under \url{https://github.com/UKAEA-IBM-STFC-Fusion-FMs/tokamark_baseline}.
  % The dataset, benchmark, documentation, and tooling are open-sourced under \url{https://github.com/UKAEA-IBM-STFC-Fusion-FMs/tokamark_baseline/tree/neurips-code#}.
\end{abstract}

\section{Introduction}

Nuclear fusion power is being explored as a potential long-term energy source with a unique combination of benefits. It offers the prospect of a carbon-neutral energy supply with abundant fuel and significant safety advantages over nuclear fission. However, commercially viable fusion demands stable, sustained operation that produces more energy than the power plant consumes---a goal made difficult by the extreme physical conditions of thermonuclear confinement under which fusion must operate \citep{donne2025beyond}.

Magnetic confinement fusion reactors confine a plasma exceeding 100 million degrees Celsius through the use of strong magnetic, as no material can withstand direct contact with such plasma \citep{Wesson,Freidberg_2007}. This environment forces all measurements to be non-invasive, only partially inferring plasma state while the underlying dynamics evolve on microsecond-to-millisecond timescales. In this work, we dedicate our attention to the problem of modeling fusion plasma dynamics in tokamaks---one of the central challenges in fusion research. This problem encompasses a wide set of predictive tasks, including plasma shape and equilibrium inference, transport and profile evolution, and forecasting of magnetohydrodynamics (MHD) activity and disruptions.

\subsection{AI for fusion plasma modeling} 

Traditional approaches to tokamak plasma modeling are rooted in well-established first-principles descriptions of magnetized plasma dynamics. These descriptions are expressed through coupled, nonlinear systems of partial differential equations, whose numerical solutions often require high-fidelity, multi-scale simulations. While such models are indispensable for predictive studies, their computational cost severely limits their routine use, both in exploring the full phenomenology of plasma behavior, and in systematically interrogating large experimental datasets. In particular, many physically relevant regimes remain difficult to characterize in detail because comprehensive parameter scans and high-fidelity simulations are prohibitively expensive.
The same computational burden complicates data-driven inference. Key parameters governing plasma behavior---such as transport coefficients, source terms, or stability-relevant profile features---are frequently unmeasured or only indirectly observable, and inferring them typically requires repeated forward simulations embedded within optimization or system identification loops. As a result, fitting models to experimental data becomes costly and brittle across operating regimes. These limitations make the direct use of first-principles solvers infeasible for real-time applications.
% , where stringent latency constraints rule out iterative or high-fidelity numerical solutions. 
% The challenge is further compounded by the experimental characteristics of tokamak sensors, operating at widely different sampling rates, spatial resolutions, and noise characteristics, producing data that is inherently heterogeneous, incomplete, multi‑rate, and noisy.

Tokamaks deploy a broad suite of heterogeneous diagnostics--magnetics, optical and X‑ray emission, microwave interferometry, and more \citep{morris2002}---mounted on or behind the reactor walls and engineered to withstand extreme heat, radiation, and electromagnetic stress. These sensors operate at widely different sampling rates, spatial resolutions, and noise characteristics, producing data that is inherently heterogeneous, incomplete, multi‑rate, and noisy.
Together, the complexity of the underlying physics and the heterogeneity of the data create an opportunity where modern AI methods can provide significant advantages and complement physics‑based modeling.
% Data‑driven models are well suited to fusing multi‑modal diagnostics, learning latent plasma representations, and capturing nonlinear, multi‑scale temporal behavior without explicit physical parameterization. 
Unlike traditional solvers, AI models can operate directly on raw measurements, handle missing or asynchronous data, and produce accurate and efficient surrogates. However, these advantages come with caveats: learned models may fail silently when operating outside their training distribution, lack guaranteed physical consistency, and are generally more difficult to interpret.

Previous work have demonstrated promising results across a variety of narrowly defined tasks, ranging from plasma shape reconstruction \citep{wan_machine-learning-based_2023-1, wai_neural_2022, rossi_potential_2023} and profile forecasting \citep{wan_experiment_2021, abbate_data-driven_2021, abbate_general_2023-1, abbate_combining_2025, char_full_2024, kit_learning_2024, wakatsuki_simultaneous_2023} to actuator optimization \citep{wang_learning_2025, wang_active_2024, yang_modeling_2020, schramm_development_2024, seo2024avoiding, seo_feedforward_2021, seo_development_2022, vega_disruption_2022, degrave_magnetic_2022, abbate_general_2023-1, orozco_neural_2022} and disruptions prediction \citep{zhang_database_2020, zhu_integrated_2023, priyanka_review_2024, rea_disruption_2018, lucas_disruptionbench_2024, montes_interpretable_2021, montes_semi-supervised_2021-1, churchill_deep_2020-1, ferreira_deep_2020, guo_disruption_2021, kates-harbeck_predicting_2019, zhu2021hybrid, de2011survey, zhu2021scenario, aymerich2022disruption}. However, most of those efforts use pipelines tailored to a small set of diagnostics, single device and single scientific objective, relying heavily on task-specific feature engineering and handcrafted labeling procedures. 
% While successful within their respective scopes, these approaches typically optimize models for individual tasks, which limits reuse across the experimental lifecycle. 
Adding to ongoing efforts \citep{Dong_2025,yang2025fusionmaelargescalepretrainedmodel}, more work is needed to move beyond isolated solutions toward broad, interoperable models capable of understanding fusion plasmas in a comprehensive way.

Inspired by the success of Foundation Models (FM), there is a growing expectation that analogous models trained on large corpora of tokamak data could learn rich, transferable plasma representations \citep{churchill2025ai} and support a wide range of downstream tasks. Although still in early stages, this paradigm suggests a path toward generalist AI systems for fusion that complement physics-based modeling and reduce the need for handcrafted pipelines \citep{churchill2025ai}.
%
% First motivation for benchmark: broader problem statement, cross-comparison, standartisation and public accesibility
% Progress toward either specialized or generalist plasma models is hampered by the absence of open, standardized benchmarks. 
Fusion datasets remain fragmented across institutions, locked behind proprietary interfaces, or stored in domain‑specific formats that are difficult for Machine Learning researchers to access or interpret \citep{strand2022fair}. Without unified task definitions, metrics, or evaluation protocols, it becomes impossible to provide a fair comparison of methods and to measure progress systematically. A benchmark is therefore needed to frame the plasma modeling problems in a broader and more structured way, enabling cross‑comparison across algorithms, reproducibility across labs, and accessibility for researchers both inside and outside the fusion community.

% Second motivation for benchmark: dedicated modeling work
% At the same time, fusion diagnostic data provide a unique set of challenges going beyond those found in standard datasets from more established areas of AI. The data is inherently multi‑modal, multi‑rate, and multi‑dimensional, with wide variations in signal quality, noise characteristics, and information content (see Figure \ref{fig:mast-data} for an illustrative example). Many diagnostics operate asynchronously, contain gaps, or suffer partial failures; others produce high‑dimensional structured outputs such as flux maps or spectrograms. This complexity demands dedicated modeling approaches that can handle missing data, fuse heterogeneous signals, different modalities, and reason across fast and slow timescales---capabilities that are difficult to stress‑test without a standardized benchmark.

Taken together, these factors create a clear need for a comprehensive benchmark suite that defines common tasks, standardizes evaluation, and provides open access to representative fusion data. Establishing such a benchmark is essential for accelerating research, enabling fair comparison of models, and ultimately advancing data-driven plasma understanding and control.

\subsection{\textbf{\bench} Benchmark Overview}

\setlength{\textfloatsep}{9pt}
\setlength{\floatsep}{3pt}
\setlength{\intextsep}{3pt}
\begin{figure}
    \centering
    \includegraphics[width=0.7\textwidth]{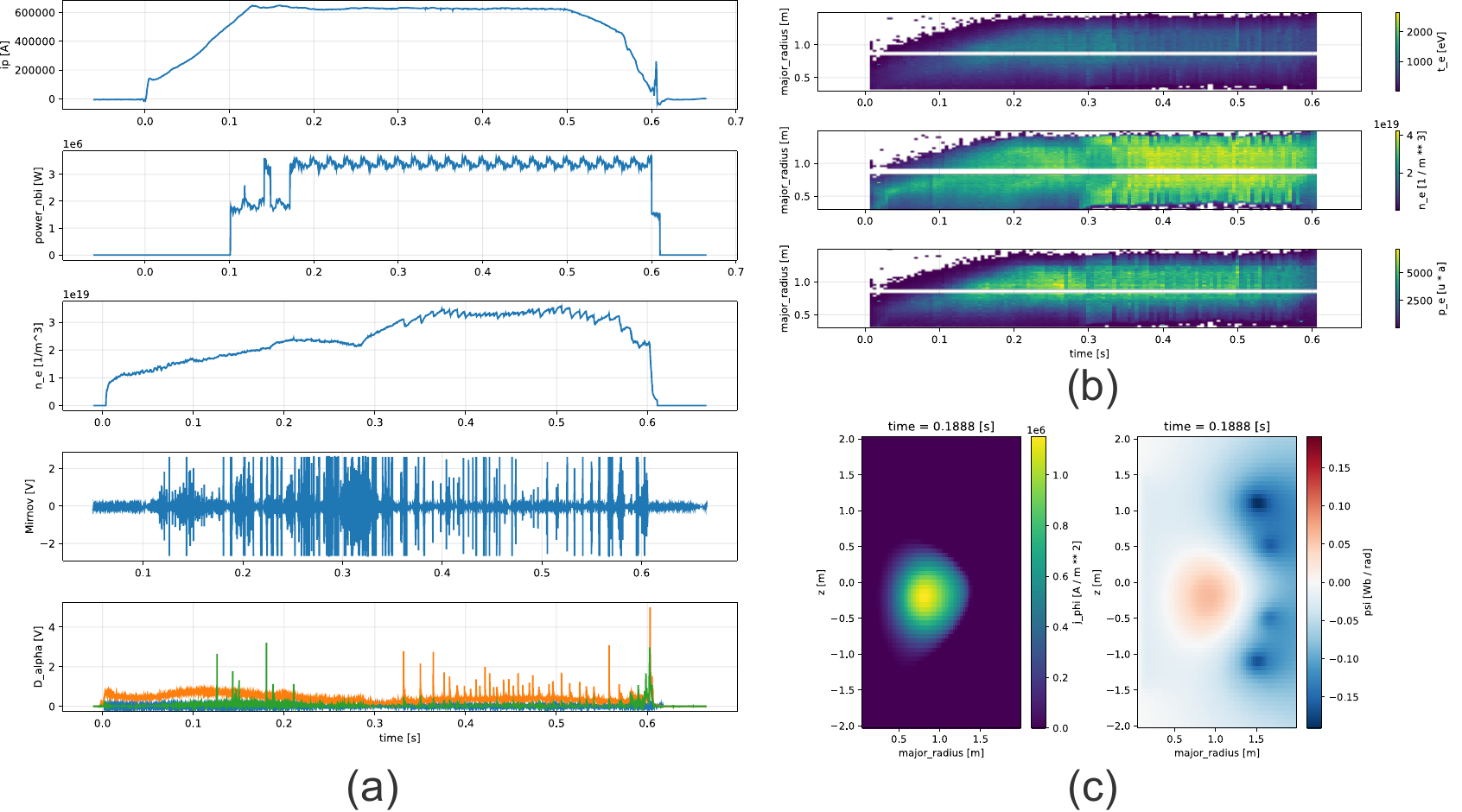}
        \captionsetup{skip=2pt}
    \caption{Examples of multi-modal signals from FAIR-MAST data: (a) time series of plasma current, line averaged density, NBI power, mirnov coils, and D$_\text{alpha}$ signals; (b) Thomson scattering profiles of electron temperature, density, and pressure; (c) maps of plasma current and poloidal magnetic flux.}
    \label{fig:mast-data}
\end{figure}

% To fill the need for comprehensive fusion plasma benchmarks, 
We introduce \bench, the first large, open benchmark for evaluating AI models trained on real fusion data. It uniquely supports a diverse range of tasks—from reconstruction to long-horizon forecasting—within a single, unified dataset, enabling consistent and comprehensive evaluation across multiple modeling objectives. This breadth of task coverage within one dataset is, to the best of our knowledge, unprecedented in fusion machine learning benchmarks.

% Data:
\textbf{Data.}
FAIR-MAST data represents the only openly available dataset of real tokamak diagnostics. Recent releases \citep{jackson2024fair, jackson_open_2025}
curate a collection of real experiments and corresponding diagnostics measurements from MAST tokamak. From FAIR-MAST, we select \TotalSignals{} signals across heterogeneous modalities, harmonize metadata, and build standardized loaders.

% Tasks:
\textbf{Tasks.}
We define a diverse suite of 14 downstream tasks organized into 4 groups, designed to probe core capabilities:
% required by AI models for fusion plasmas
(i) representation learning; (ii) temporal reasoning across fast and slow timescales; (iii) robustness to incomplete state information; and (iv) generalization across operating regimes. 
Rather than optimizing for a single downstream objective, the tasks span a cascade of physical processes---from fast magnetic response to slower, transport-driven evolution, and long-horizon precursors of MHD activity---while remaining closely aligned with routine experimental workflows. Wherever possible, the tasks minimize reliance on expert-labeled targets, supporting self-supervised and weakly supervised formulations, and enabling systematic evaluation of transferable plasma representations. 

% Evaluation
\textbf{Evaluation.}
We introduce a hierarchical evaluation protocol aligned with the structure of FAIR-MAST and with scientific objectives. 
This hierarchy assesses both low-level prediction quality and high-level scientific utility. 
For specialized models (single-task training), the hierarchy yields granular, signal-level diagnostics; for generalist models (FM-style pretrained and then fine-tuned across tasks), it provides modular comparisons across tasks and broad scientific objectives.

% Baseline
\textbf{Baseline.}
Finally, we provide a strong yet accessible baseline: two simple statistical models (\textbf{Persistence} and \textbf{Mean}), and a \emph{multi-branch convolutional encoder--decoder} architecture with two variants (\textbf{CNN} and \textbf{CNN+LSTM}) inspired by the previous works \citep{seo2023multimodal, seo2024avoiding}, trained independently for each task, and largely expanded to ingest heterogeneous inputs and outputs.

The following list summarizes our contributions:
\begin{itemize}[topsep=0pt, partopsep=0pt, parsep=0pt, itemsep=0pt]
    \item \textbf{Benchmark design.} We define 14 tasks organized into four
    groups, with a hierarchical evaluation protocol, standardized windowing, and error metrics.
    \item \textbf{Data packaging.} We freeze a stable subset of FAIR-MAST data, standardize metadata and units, and release it for reproducibility and long-time compatibility.
    \item \textbf{Tools and API.} We provide a Python package for task-specific
    data loading, processing and batching, alignment utilities,
    and evaluation logic, integrated with the PyTorch stack.
    \item \textbf{Baseline models.} We release both naive and advanced models
    % multi-branch convolutional encoder--decoder baseline
    with configurations and training scripts, establishing reproducible baselines across all tasks.
\end{itemize}

\begin{table*}[t]
\centering
\caption{Signals taxonomy for \bench.}
\label{tab:taxonomy}
\scriptsize

\setlength{\tabcolsep}{4pt}
\setlength{\aboverulesep}{0.5ex}  % space above rules
\setlength{\belowrulesep}{0.5ex}  % space below rules

\begin{tabular}{lllll}
\toprule
\textbf{Category} & \textbf{Subcategory/Signals} & \textbf{Origin} & \textbf{Frequency} & \textbf{Modality} \\
\midrule
\textbf{Magnetics} & Flux loops, pickup coils; saddle coils & Diagnostic & 5 kHz; 50 kHz & Profile \\
\textbf{Kinetics} & Thomson scattering; interferometer & Diagnostic & 0.2 kHz; 4 kHz & Profile; time series \\
\textbf{Radiatives} & D$_\text{alpha}$, soft X-ray & Diagnostic & 50 kHz & Profile \\
\textbf{Fast magnetics} & Mirnov coils & Diagnostic & 500 kHz & Profile \\
\textbf{Currents} & Poloidal field coil currents; solenoid and plasma current & Diagnostic & 4 kHz & Profile; time series \\
\textbf{Voltages} & Poloidal field coil voltages & Actuator & 4 kHz & Profile \\
\textbf{References} & Reference plasma current, reference plasma density & Actuator & 4 kHz & Time series \\
\textbf{Fueling} & NBI power, gas puffing & Actuator & 4 kHz & Time series \\
% \textbf{Plasma Shape} & Shape parameters; plasma boundary & Derived & 10--200 Hz (recon cadence) & 1D contour (1D$\times$t) \\
\textbf{Equilibrium} & Shape parameters, $J_{\text{tor}}$ metrics; flux map & Derived & 0.2 kHz & Time series; profile; video \\
\bottomrule
\end{tabular}
\end{table*}

% ==================================================================================s
\section{Preliminaries}
% ==================================================================================

To ground \bench\ benchmark in the reality of complex heterogeneous fusion data, we first introduce a signal taxonomy that organizes these measurements by function and modality, which allows us to provide a consistent vocabulary for describing inputs, targets, and tasks in the benchmark.

% ----------------------------------------------------------------------------------
\subsection{Data Taxonomy}
% ----------------------------------------------------------------------------------

MAST is a spherical tokamak located in Culham, Oxfordshire (UK) that was operated by UKAEA and EURATOM from 1999 to 2013 \citep{sykes2001first, counsell2005overview, meyer2009overview}. MAST, and tokamaks in general, operates in short experimental cycles known as \emph{discharges} or \textbf{shots}. The length of those cycles depends on the device size, and in the case of MAST they typically last around 2–3 seconds. Over its operational lifetime, MAST produced more than 30,000 shots, with each shot containing diagnostic signals measuring various properties of the plasma, including the magnetic field, plasma temperature, shape parameters, and applied heating, to name a few. Recent works \citep{jackson2024fair, jackson_open_2025} have created an open dataset of diagnostic data from a subset of the history of MAST. FAIR-MAST data contains \DatasetSize\ shots from the last five experimental campaigns on MAST. In this work, we utilized a total of \TotalSignals\ signals from FAIR-MAST data for the design of our benchmark tasks. We organize all the selected signals along several complementary axes, as summarized in Table~\ref{tab:taxonomy}. 

First, we group signals by \emph{category}, which encodes their physical semantics and how they are used by MAST workflows. This information corresponds to the first two columns of Table~\ref{tab:taxonomy}. Note that the second column contains both individual signals and \emph{subcategories}. These intermediate groupings are introduced for brevity: they consolidate multiple signals that are typically used together (at least within the proposed benchmark) and share the same structural and functional properties. For example, the \emph{shape parameters} subcategory includes several attributes jointly describing plasma geometry. A full list of the signals used in this work is provided in the appendix in Table~\ref{tab:full_list_of_signals}.

Second, we distinguish signals by their \emph{origin}, for which we define three classes: (i) \textbf{diagnostics}, corresponding to direct hardware measurements; (ii) \textbf{actuators}, representing controllable machine parameters used to steer plasma behavior; and (iii) \textbf{derived} signals, produced by reconstruction pipelines such as EFIT \citep{APPEL20181} (e.g., shape parameters or flux maps).

The third axis in Table~\ref{tab:taxonomy} is \emph{frequency}, reflecting the wide range of sampling rates across diagnostics, spanning from 0.2~kHz up to 500~kHz. Finally, the fourth axis is \emph{modality}, describing the structural form of each signal: (i) \textbf{time series}, represented as 1D tensors (scalar value over time); (ii) \textbf{profiles}, represented as 2D tensors (vector over time); and (iii) \textbf{videos}, represented as 3D tensors (image over time). These latter two axes are particularly important for guiding the design of AI model architectures and loss functions.

% ----------------------------------------------------------------------------------
\subsection{Structural Taxonomy of Tasks}
% ----------------------------------------------------------------------------------

The downstream tasks in \bench\ share a common structural formulation designed to reflect the online, window-based nature of plasma control and forecasting in tokamak experiments. Rather than operating on full-shot signals (as is often done in offline post-discharge analysis), our tasks are defined using \textbf{input window} and \textbf{output window} anchored at a reference time point.

Each task consists of one or more \emph{input signals}, and one or more \emph{output signals} also referred to as targets. Inputs typically include both diagnostic signals (e.g., magnetics, radiatives, kinetics) and actuator signals (e.g., voltages, fueling), while outputs include either diagnostic signals or derived quantities like equilibrium reconstructions. The alignment between input and output windows naturally induces
families of modeling objectives:
\begin{itemize}[topsep=0pt, partopsep=0pt, parsep=0pt, itemsep=0pt]
    \item \textbf{Reconstruction.}
    Given a set of diagnostic signals $A$ over the interval $[t_0-\Delta_{input}, t_0]$, the goal is to reconstruct a related set of signals $B$ over the \emph{same} interval.
    % This corresponds to instantaneous or near-instantaneous inference, often used
    % for equilibrium reconstruction or diagnostic fusion.

    \item \textbf{Autoregressive (AR) Forecasting.}
    Given diagnostic signals $A$ over $[t_0-\Delta_{input}, t_0]$ together with actuator trajectories over $[t_0-\Delta_{input}, t_0+\Delta_{output}]$, predict future values of the same diagnostic signals $A$ over $[t_0, t_0+\Delta_{output}]$. 
    % This is the dominant structure for short-horizon plasma dynamics forecasting.

    \item \textbf{Reconstructive (RC) Forecasting.}
    Using diagnostic signals $A$ over $[t_0-\Delta_{input}, t_0]$ and actuators over $[t_0-\Delta_{input}, t_0+\Delta_{output}]$, forecast a set of related outputs $B$ over $[t_0, t_0+\Delta_{output}]$ where $B$ may contain signals as in $A$.
    % This formulation generalizes forecasting to cases where future predictions include additional derived quantities beyond the original inputs.
\end{itemize}
We further distinguish tasks by their temporal dependency structure: \textbf{Markovian} tasks require only a short input window to make forecasts, reflecting fast dynamics, whereas \textbf{Non-Markovian (NM)} tasks require substantially longer input histories even for short-term predictions.

% ----------------------------------------------------------------------------------
\subsection{Data-driven Challenges}
% ----------------------------------------------------------------------------------

% Working with fusion data is challenging \citep{spangher_position_nodate}. Physical phenomena of a plasma operate over a wide range of time scales and are measured with heterogeneous diagnostic systems. Specifically, there are three core challenges with working with fusion diagnostic data:

To complete the description of FAIR-MAST data, we highlight three core challenges in resolving tasks in \bench, arising from the 
% heterogeneous, multi-instrument nature of tokamak diagnostics and the 
operational realities of large fusion experiments.

\textbf{Multi-fidelity.} 
Diagnostic systems operate at varying sampling rates.
% , ranging from a few hundred to several hundred kilohertz. 
Resampling to a common high-frequency time base is computationally expensive and often unnecessary, while down-sampling to the lowest frequency discards information critical for resolving fast plasma phenomena. Effective models must therefore integrate and represent multi-rate temporal data without losing fidelity.
    
\textbf{Multi-modality.}
Signals differ not only in physical meaning, units, and numerical ranges, but also in structural forms of the corresponding data tensors. Their dimensionality can be from 1 to 3, with the number of channel in each non-temporal dimension varying from 1 up to 170. 
% AI systems must be able to fuse these modalities into coherent representations suitable for downstream tasks.
 
\textbf{Missing data.}
As with any experimental dataset, missing information is common as demonstrated by the per signal missing-data statistics reported in Appendix~\ref{sec:stat_summ}. Entire signals may be absent for a given shot due to hardware issues. Individual signals may also contain missing time segments, for example due to limited acquisition windows or diagnostic failures. Naively discarding shots or windows with missing components wastes valuable examples and can introduce distributional bias. Robust approaches must therefore handle incomplete signals and irregular temporal coverage.

% ==================================================================================
\section{\bench: A MAST Benchmark}
% ==================================================================================

We present the main components of \bench: tasks, data preparation, and evaluation protocol.

% ----------------------------------------------------------------------------------
\subsection{Downstream Tasks}
% ----------------------------------------------------------------------------------

\begin{table*}[t]
\centering
\caption{Summary of groups and tasks for \bench.}
\label{tab:downstream-tasks}
\scriptsize

\renewcommand{\arraystretch}{0.85}
\setlength{\aboverulesep}{0.5ex}  % space above rules
\setlength{\belowrulesep}{0.5ex}  % space below rules
\setlength{\tabcolsep}{4pt}

\begin{tabular}{l l l l l l l l}
\toprule
\textbf{} &
\textbf{Task} &
\textbf{Input Diagnostics} &
\textbf{Input Actuators} &
\textbf{Outputs} &
\textbf{Type} &
% \makecell[l]{\textbf{Input} \\ \textbf{Window} / \textbf{Output Window}} \\
% \makecell[l]{\textbf{Input} \\ \textbf{Window}} & \makecell[l]{\textbf{Output} \\ \textbf{Window}} \\
\makecell[l]{\textbf{Input}} & \makecell[l]{\textbf{Output}} \\
\midrule

% ---------------- GROUP 1 ----------------
\multirow{3}{*}{\rotatebox{90}{\textbf{Group 1}}} &
1-1 &
\makecell[l]{Magnetics, Currents} &
-- &
\makecell[l]{Shape parameters, \\$J_{\text{tor}}$ metrics} &
Reconstruction &
5 ms & 5 ms \\

& 1-2 &
Same as 1-1 &
-- &
Plasma boundary &
Reconstruction &
5 ms & 5 ms \\

& 1-3 &
Same as 1-1 &
-- &
Flux map &
Reconstruction &
5 ms & 5 ms \\
\midrule

% ---------------- GROUP 2 ----------------
\multirow{3}{*}{\rotatebox{90}{\textbf{Group 2}}} &
2-1 &
\makecell[l]{Magnetics, Currents} &
\makecell[l]{Voltages, \\ NBI Power} &
\makecell[l]{Currents, $J_{\text{tor}}$ metrics,\\ Shape parameters} &
\makecell[l]{RC Forecasting} &
5 ms & 25 ms \\

& 2-2 &
Same as 2-1 &
Same as 2-1 &
Plasma boundary &
\makecell[l]{RC Forecasting} &
5 ms & 25 ms \\

& 2-3 &
Same as 2-1 &
Same as 2-1 &
Flux map &
\makecell[l]{RC Forecasting} &
5 ms & 25 ms \\
\midrule

% ---------------- GROUP 3 ----------------
\multirow{3}{*}{\rotatebox{90}{\textbf{Group 3}}} &
3-1 &
\makecell[l]{Thomson scattering} &
\makecell[l]{References, Fueling} &
Thomson scattering &
\makecell[l]{AR Forecasting} &
5 ms & 50 ms \\

& 3-2 &
\makecell[l]{Thomson scattering, Radiatives} &
Same as 3-1 &
Radiatives &
\makecell[l]{AR Forecasting} &
5 ms & 50 ms \\

& 3-3 &
\makecell[l]{Magnetics, Currents, \\  Interferometer} &
Same as 3-1 &
\makecell[l]{Thomson scattering, \\ $J_{\text{tor}}$ metrics} &
\makecell[l]{RC Forecasting, NM} &
history & 5 ms \\

\midrule

% ---------------- GROUP 4 ----------------
\multirow{5}{*}{\rotatebox{90}{\textbf{Group 4}}} &
4-1 &
\makecell[l]{Magnetics, Mirnov coils, \\ Currents, Radiatives, \\Interferometer} &
References, Fueling &
Soft X-ray &
\makecell[l]{AR Forecasting, NM} &
history & 100 ms \\

& 4-2 &
\makecell[l]{Magnetics, Mirnov coils, \\ Currents, Radiatives, Kinetics} &
Same as 4-1 &
Soft X-ray &
\makecell[l]{AR Forecasting, NM} &
history & 100 ms \\

& 4-3 &
Same as 4-1 &
Same as 4-1 &
Shape parameters &
\makecell[l]{RC Forecasting, NM} &
history & 100 ms \\

& 4-4 &
Same as 4-1 &
Same as 4-1 &
Plasma current &
\makecell[l]{AR Forecasting, NM} &
history & 100 ms \\

& 4-5 &
Same as 4-1 &
Same as 4-1 &
Mirnov diagnostics &
\makecell[l]{AR Forecasting, NM} &
history & 100 ms \\
\bottomrule

\end{tabular}
\end{table*}

In \bench, we assembled a set of 14 tasks. We divided them into 4 groups representing some of the major modeling challenges arising in real-world fusion experiments. The 4 groups include: instantaneous reconstruction, short‑term magnetics dynamics, slow transport‑driven profile evolution, and long‑range forecasting of MHD activity. Table~\ref{tab:downstream-tasks} summarizes definitions of tasks and groups.

% ..................................................................................
\subsubsection{Group 1: Instantaneous Equilibrium Reconstruction}
% ..................................................................................

Group 1 defines a suite of reconstruction problems where the objective is to infer plasma equilibrium---its shape, boundaries, and various properties---from magnetic diagnostics and coil currents. The targets of this group span multiple modalities from scalar parameters (Task 1-1) to full plasma contour (Task 1-2) and two-dimensional representations of the poloidal magnetic flux (Task 1-3). 

Equilibrium reconstruction is performed routinely after every plasma discharge, traditionally by solving an inverse boundary-value problem governed by the Grad–Shafranov equation that describes the balance of magnetohydrodynamic force in axisymmetric plasmas. Solvers such as EFIT++ \citep{APPEL20181} require iterative convergence and parameterized profile assumptions, limiting real-time deployment. Group 1 evaluate whether models can \textbf{infer equilibrium} directly from raw diagnostics, providing fast, numerics-free surrogates for both offline feedback and real-time control.

% ..................................................................................
\subsubsection{Group 2: Short-Term Magnetics Dynamics}
% ..................................................................................

Group 2 focuses on short-timescale forecasting of magnetic signals, coil currents, and equilibrium evolution in response to actuator commands. While conceptually similar to Group 1, these tasks shift from static reconstruction to sequence-to-sequence prediction, with actuator signals becoming essential inputs. Task complexity increases across the group, ranging from scalar forecasts such as plasma current evolution to joint prediction of 2D equilibrium geometry over short horizons.

At these timescales, dynamics are dominated by inductive coupling between active coils, passive structures, and plasma current, as well as the plasma’s local response to perturbations. These effects govern plasma position, shape, and current control during a discharge and are typically modeled offline. Group 2 probes whether models can learn an effective description of the \textbf{plasma response to control actions} via the coupled magnetic dynamics of the active coils and vessel. This would directly target capabilities required for closed-loop control, scenario planning, and digital twin applications.

% ..................................................................................
\subsubsection{Group 3: Kinetic Profile Dynamics}
% ..................................................................................

Group 3 tasks focus on modeling the temporal evolution of kinetic profiles---primarily, electron density and temperature---and on forecasting diagnostic signals associated with confinement-mode transitions. The three tasks in this group include short-horizon forecasting (Task 3-1 and Task 3-2) and pseudo-real-time reconstruction using partial, multi-rate diagnostic inputs (Task 3-3).

Profile evolution is governed by transport physics acting on particle and energy balance, introducing slower characteristic timescales and intrinsic memory effects compared to magnetic dynamics. These processes control energy confinement and overall plasma performance and are central to understanding transport, optimizing heating schemes, and predicting access to high-performance regimes. In practice, profile measurements are often sparse, delayed, or unavailable in real time. Group 3 assesses AI models' ability to \textbf{infer latent plasma state} from incomplete diagnostic information. This would support applications ranging from post-shot analysis, to integrated scenario design and real-time performance monitoring and control.

% ..................................................................................
\subsubsection{Group 4: Long-Range Forecasting of MHD activity}
% ..................................................................................

Group 4 comprises long-horizon forecasting of thermal quenches (Task 4-1 and Task 4-2), vertical displacement events (Task 4-3), current quenches (Task 4-4), and MHD activity {leading to locked modes} (Task 4-5). These tasks are focused on \textbf{detection of plasma instability} and disruptions precursors that emerge across multiple diagnostic modalities and require processing high-frequency signals---such as magnetics, radiative diagnostics, and MHD signatures---over extended temporal windows. Notably, Task 4-5 operates in the Fourier domain, predicting spectral features of Mirnov signals rather than time-domain waveforms, for a more robust representation of MHD activity.

These instabilities are tightly linked to the evolving equilibrium and current profiles, even when the nonlinear dynamics of the instability itself are difficult to model explicitly. Disruption avoidance and mitigation are critical operational requirements, directly affecting machine lifespan. In practice, early-warning systems rely on detecting subtle precursors distributed across multiple diagnostics and evolving over extended time windows. Successful performance in Group 4 tasks would demonstrate the ability to integrate long-range temporal context and multi-modal information, allowing the models to anticipate loss-of-control events, a key requirement for safe and reliable fusion operation.

\subsubsection{Input and output window length}
For fast magnetic and actuator dynamics (Group 2, Tasks 3-1 and 3-2), the observed diagnostics provide a sufficient description of the system state, such that short input windows are often adequate to predict near-term evolution. 
In contrast, profile evolution (Task 3-3), confinement transitions, and MHD activity (Group 4) depend on latent plasma state variables
% ---such as current and pressure profiles---
that are only partially and indirectly observed. For these processes, accurate prediction requires integrating information over extended time intervals. By explicitly varying the temporal context required across tasks, the benchmark probes a model’s ability to learn both short-timescale system dynamics and long-range temporal dependencies arising from unobserved physics.

% ----------------------------------------------------------------------------------
\subsection{Data Preparation}
% ----------------------------------------------------------------------------------

We take all \DatasetSize\ available shots from FAIR-MAST data and extract the data for \TotalSignals\ signals required by our task definitions. Before being used for model training and evaluation, these signals undergo a standardized preprocessing pipeline to ensure consistent formatting and alignment.

\textbf{Data filtering.} The dataset was filtered to ensure consistency and physical validity by removing non-physical or burnt values, sentinel entries, and retaining only converged equilibrium reconstruction. We note that signals are used at their original sampling rates, and no resampling or imputation is performed, with evaluation performed using masked metrics on available ground truth.

\textbf{Data split.} The benchmark dataset is divided into disjoint \textit{training}, \textit{validation}, and \textit{test} subsets. To avoid information leakage across sets, the split is performed at the shot level representing independent experiments. Two splitting strategies are considered. First, a random sampling technique using an 80-10-10 ratio is employed. Second, a temporal split is introduced to assess generalization across campaigns, where the most recent two experimental campaigns are held out as the test set. All hyperparameter tuning is conducted exclusively on the training and validation sets. The number of shots used for each task under both splitting strategies is provided in Table~\ref{tab:task_random_temporal}.

\textbf{Window segmentation.} Shot-level signals data is segmented into input and output windows. The lengths of those windows are task-dependent as specified in Table 2, and a stride of 0.001s is used. The chosen stride is a compromise between heterogeneous diagnostic sampling rates and the relevant physical timescales—capturing fast dynamics: it captures the fast evolution of magnetics and control signals while remaining compatible with slower diagnostics. Each input window is paired with its corresponding output window and treated as an independent window-level sample.

% ----------------------------------------------------------------------------------
\subsection{Benchmark Evaluation}
% ----------------------------------------------------------------------------------

We introduce a hierarchical evaluation protocol which explicitly separates three levels of the hierarchy: \textbf{signals} (individual physical quantities), \textbf{tasks} (well-defined scientific goals), and \textbf{groups} (broader physical objectives). This provides both, signal-level insights that help diagnose which physics regimes are captured, and higher-level scientific utility assessment. To respect the hierarchy, the evaluation aggregates errors according to the following progression:
{
\small
\setlength{\abovedisplayskip}{1pt}
\setlength{\belowdisplayskip}{1pt}
\[
\text{samples} \;\rightarrow\; \text{windows} \;\rightarrow\; \text{signals}
\;\rightarrow\; \text{tasks} \;\rightarrow\; \text{shots}.
\]
}

\textbf{Samples} are the atomic level of data. For a given task and shot, each data sample is denoted as $y_{k,i,j}$ and corresponds to a particular sample $j$ of the flattened data from window $i$ and signal $k$. In the following, $y_{k,i,j}$ denotes the ground truth value and $\hat y_{k,i,j}$ the corresponding model prediction. 
\textbf{Windows} are containers of equal per-signal size storing $N_{k}$ samples. We compute a window-level \ac{rmse} as:
{\small
\abovedisplayskip=1pt
\belowdisplayskip=1pt
\begin{equation}
\mathrm{RMSE}_{k,i} = \sqrt{
    \frac{1}{N_{k}}
    \sum_{j=1}^{N_{k}}
    \left( y_{k,i,j} - \hat y_{k,i,j} \right)^2} \quad
\end{equation}
}

% For each \textbf{signal} $k$ containing $M_k$ windows, we aggregate window errors and normalize by the global empirical standard deviation $\sigma_{k}$:
For each \textbf{signal} $k$ for a given task and shot and containing $M_k$ windows, we aggregate all window errors via the single-shot signal \ac{rmse} and normalize it with the global empirical standard deviation $\sigma_{k}$ computed for signal $k$ across all evaluation shots. This yields a dimensionless quantity that expresses prediction error relative to the natural variability of the target, making the metric comparable across signals.
Concerning \textbf{tasks}, task-level errors $\tilde{e}_{t}$ combine $K_t$ normalized output signal errors into a single score, quantifying model performance on the scientific objective as a whole:
{
\small
\abovedisplayskip=0.7pt
\belowdisplayskip=0.7pt
\begin{equation}
% \mathrm{RMSE}_{k} = 
% \sqrt{
%     \frac{1}{M_{k}}
%     \sum_{i=1}^{M_{k}} \mathrm{RMSE}_{k,i}^2}, \quad
\mathrm{RMSE}_{k} = \sqrt{
    \frac{1}{M_{k}}
    \sum_{i=1}^{M_{k}} \mathrm{RMSE}_{k,i}^2}, \quad
    \mathrm{NRMSE}_{k} = \frac{\mathrm{RMSE}_{k}}{\sigma_{k}}, \quad
    \tilde{e}_{t} = \frac{1}{K_t} \sum_{k=1}^{K_t} \mathrm{NRMSE}_{k}.
\end{equation}
}
% which yields a dimensionless quantity that expresses prediction error relative to the natural variability of the target, making the metric comparable across signals. 
% An error of $\mathrm{NRMSE}_{k} = 1$ corresponds to a model no better than approximating the signal by its mean, while $\mathrm{NRMSE}_{k} < 1$ indicates predictive value. 

% Concerning \textbf{tasks}, task-level errors $\tilde{e}_{t}$ combine $K_t$ normalized output signal errors into a single score using a uniform mean, allowing one to quantify model performance on the scientific objective as a whole:
% {
% \abovedisplayskip=1pt
% \belowdisplayskip=1pt
% \begin{equation}
%     \tilde{e}_{t} = \frac{1}{K_t} \sum_{k=1}^{K_t} \mathrm{NRMSE}_{k}.
% \end{equation}
% }
% \textbf{Groups} combine task-level errors of the related tasks. Assuming group $g$ contains $T_g$ tasks, group-level error is computed as:
% \begin{equation}
%     \hat{e}_{g} = \frac{1}{T_g} \sum_{t=1}^{T_g} \tilde{e}_{t}.
% \end{equation}

\textbf{Shots} represent independent experimental realizations, and therefore provide the final step of aggregation.
% In the \bench{} we report three types of shot-level averaged scores that reflect the structure of the hierarchy: i) per-signal score, i) per-task score and ii) per-group score; correspondingly defined as follows
We report signal-level and task-level errors as NRMSE based aggregated across shots quantities correspondingly defined as:
{\small
\abovedisplayskip=0.7pt
\belowdisplayskip=0.7pt
\begin{align}
    % \tilde{e}_{t} = \frac{1}{K_t} \sum_{k=1}^{K_t} \mathrm{NRMSE}_{k}, \quad
    \text{Signal}_{\mathrm{NRMSE}} =
    \frac{1}{S} \sum_{s=1}^{S} \mathrm{NRMSE}_{k(s)}, \quad
    \text{Task}_{\mathrm{NRMSE}} =
    \frac{1}{S} \sum_{s=1}^{S} \tilde{e}_{t(s)}.
    % \mathrm{GroupScore} &=
    % \frac{1}{K} \sum_{s=1}^{K} \hat{e}_{s,g}
\end{align}
}

Here, $S$ is the total number of shots in the test set used for evaluation.
Finally, the $\text{Group}_{\mathrm{NRMSE}}$ score is taken as an average of the corresponding $\text{Task}_{\mathrm{NRMSE}}$ scores.

% ==================================================================================
\section{Experiments with Baseline Models}
\label{sec:baseline}
% ==================================================================================

\begin{figure*}
    \centering
    \includegraphics[width=0.9\textwidth]{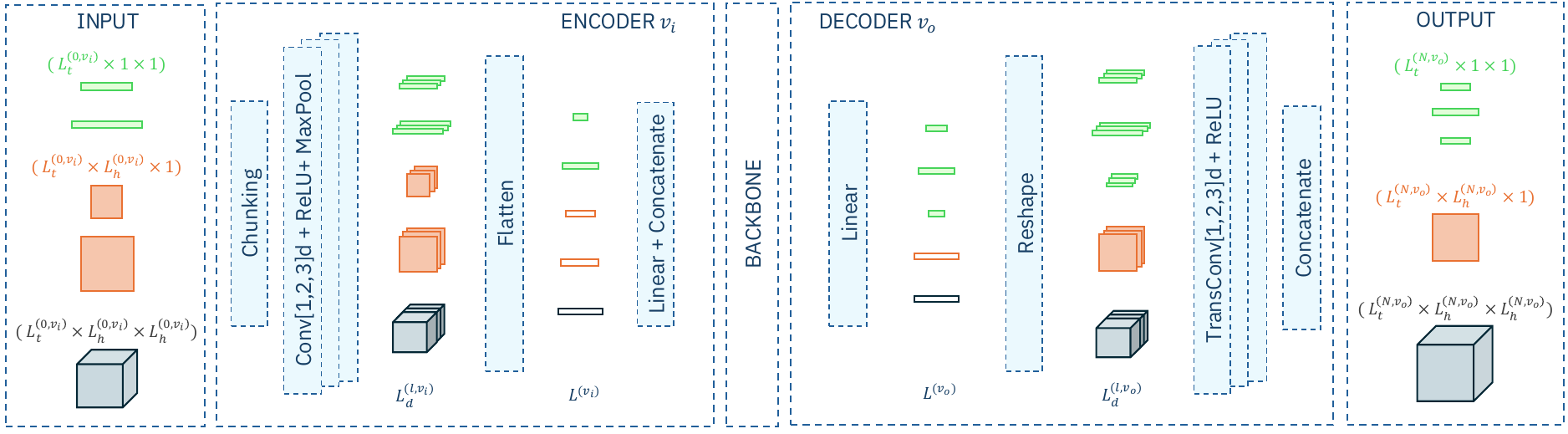}
    \captionsetup{skip=0pt}
    \caption{Multi-branch convolutional encoder–decoder architecture.}
    \label{fig:cnn-wip}
\end{figure*}

The nature of diagnostics poses significant architectural challenges, motivating the development of a dedicated baseline. Our baseline is a multi-branch convolutional architecture in which each input modality is processed by a dedicated encoder \citep{seo2023multimodal} and each target variable is generated by a corresponding decoder, all connected through a shared latent representation, allowing the model to handle inputs and outputs of varying dimensionality and temporal resolution (see Figure~\ref{fig:cnn-wip}).

% ----------------------------------------------------------------------------------
\subsection{Description of Baseline Models}
% ----------------------------------------------------------------------------------

% ..................................................................................
\subsubsection{Naive Models: Persistence and Mean Models}
% ..................................................................................

We include two basic predictors as simple statistical baselines: the \textbf{Persistence Model}, which performs AR Forecasting for selected tasks by propagating initial conditions; and the \textbf{Mean Model}, which performs prediction for all tasks using mean values.

% ..................................................................................
\subsubsection{Advanced Models: Multi-branch Convolutional Encoder–Decoder Models}
% ..................................................................................

\setlength{\textfloatsep}{3pt}
\setlength{\floatsep}{3pt}
\setlength{\intextsep}{3pt}

The proposed framework consists of three components: window-based preprocessing, modality-specific encoder–decoder networks, and a latent fusion backbone. 

Input and actuator signals are partitioned into non-overlapping 5 ms temporal chunks and standardized using z-score normalization, with missing values set to zero to obtain dense inputs. Each modality is processed by a dedicated convolutional encoder, consisting of $N$ convolutional blocks (convolution, batch normalization, ReLU, and max-pooling) that progressively reduce resolution to dimensionality $D$ while increasing feature dimensionality. Decoders symmetrically mirror encoders using transposed convolutions with cropping for output alignment. Latent embeddings from all encoder branches are concatenated and passed through a shared fusion backbone. We consider two variants: an MLP-based backbone (\textbf{CNN}) with fully connected layers and dropout, and an LSTM-based backbone (\textbf{CNN+LSTM}) that models temporal dependencies by processing embeddings as a sequence. Further details are provided in Appendix~\ref{sec:baselines}.

% ----------------------------------------------------------------------------------
\subsection{Experimental Results}
% ----------------------------------------------------------------------------------

% \begin{table}[t]
\begin{table}[H]
\centering
\caption{$\textbf{Task/Group}_{\mathrm{\textbf{NRMSE}}}$ for \bench~ tasks under Random and Temporal splitting.}
\scriptsize
\renewcommand{\arraystretch}{0.9}
\setlength{\aboverulesep}{0.5ex}
\setlength{\belowrulesep}{0.5ex}

\begin{tabular}{l c cccc cccc}
\toprule
& \multicolumn{4}{c}{\textbf{Random splitting}} & \multicolumn{4}{c}{\textbf{Temporal splitting}} \\
\cmidrule(lr){2-5} \cmidrule(lr){6-9}
& Mean & Persistence & CNN & CNN+LSTM & Mean & Persistence & CNN & CNN+LSTM \\
\midrule

Task 1-1 & 0.7566 & --- & \textbf{0.2091} & 0.2215 & 0.8793 & --- & \textbf{0.6192} & 0.6232 \\
Task 1-2 & 0.9984 & --- & \textbf{0.0473} & 0.0510 & 1.0003 & --- & 0.1349 & \textbf{0.1254} \\
Task 1-3 & 0.9267 & --- & \textbf{0.1481} & 0.1591 & 0.9509 & --- & 0.5016 & \textbf{0.2567} \\
\midrule
\emph{Group 1} & 0.8939 & --- & \textbf{0.1348} & 0.1439 & 0.9435 & --- & 0.4186 & \textbf{0.3351} \\
\midrule

Task 2-1 & 0.8174 & --- & 0.2751 & \textbf{0.2197} & 0.8966 & --- & 1.7570 & \textbf{0.5597} \\
Task 2-2 & 0.9914 & --- & 0.0662 & \textbf{0.0568} & 0.9924 & --- & 0.1339 & \textbf{0.1251} \\
Task 2-3 & 0.9230 & --- & 0.1358 & \textbf{0.1263} & 0.9289 & --- & 0.2296 & \textbf{0.1937} \\
\midrule
\emph{Group 2} & 0.9106 & --- & 0.1590 & \textbf{0.1343} & 0.9393 & --- & 0.7068 & \textbf{0.2928} \\
\midrule

Task 3-1 & 0.9837 & 0.5364 & 0.3962 & \textbf{0.3743} & 0.9686 & 0.5352 & 0.3807 & \textbf{0.3581} \\
Task 3-2 & 0.6297 & 0.3651 & \textbf{0.3309} & 0.3353 & 0.7701 & \textbf{0.2685} & 0.9110 & 0.9275 \\
Task 3-3 & 0.8412 & ---    & 0.3859 & \textbf{0.3187} & 0.9859 & ---    & \textbf{0.7520} & 1.4718 \\
\midrule
\emph{Group 3} & 0.8182 & --- & 0.3710 & \textbf{0.3428} & 0.9082 & --- & \textbf{0.6812} & 0.9191 \\
\midrule

Task 4-1 & 0.5114 & 0.3247 & 0.2751 & \textbf{0.2274} & 0.6928 & \textbf{0.1597} & 1.0970 & 1.1788 \\
Task 4-2 & 0.5114 & 0.3247 & 0.2737 & \textbf{0.2286} & 0.6928 & \textbf{0.1597} & 3.0603 & 1.1445 \\
Task 4-3 & 0.7313 & ---    & 0.1291 & \textbf{0.1179} & 0.8633 & --- & 1.3249 & \textbf{0.8198} \\
Task 4-4 & 1.0271 & 0.7178 & 0.5299 & \textbf{0.4039} & 0.9239 & 0.6390 & 0.5786 & \textbf{0.4954} \\
Task 4-5 & 0.6533 & 0.8701 & 0.6507 & \textbf{0.6488} & 0.8196 & 1.0235 & 1.0179 & \textbf{0.7309} \\
\midrule
\emph{Group 4} & 0.6869 & --- & 0.3717 & \textbf{0.3255} & \textbf{0.7985} & --- & 1.4157 & 0.8062 \\

\bottomrule
\end{tabular}

\label{tab:task_score}
\end{table}

The results of the baseline models evaluation on \bench\ tasks are presented in Table~\ref{tab:task_score}, which demonstrate a clear distinction between tasks complexities: the CNN models perform well for equilibrium reconstruction (Group 1) and magnetics dynamics (Group 2) tasks, with the group-level scores yielding the lowest NRMSE values and the most consistent improvements of the CNN+LSTM variant over the naive models. These results degrade for profile dynamics (Group 3) and MHD activities (Group 4), confirming these are more difficult tasks. Notably, even within the same group, tasks scores vary substantially. For instance, plasma boundary reconstruction and forecasting tasks Task 1-2 and Task 2-2 are resolved much better than their counterparts within the respective groups.

Overall, deep learning models substantially outperform naive baselines under random splitting, with CNN+LSTM achieving the lowest NRMSE in the majority of tasks. However, temporal splitting reveals a markedly different picture, exposing generalization failures that random splitting conceals. Indeed, the gap between strategies is modest for Groups 1 and 2 but widens substantially for Groups 3 and 4. Notably, soft x-rays forecasting (Task 4-1 and Task 4-2) is by far the worst and exceeds unity, suggesting the corresponding signals are poorly constrained or inadequately represented. This behavior might indicate the presence of a distribution shift between training and test sets, potentially due to sensor recalibration or tuning between experimental campaigns.

Finally, we provide signal level error metrics in Appendix~\ref{sec:evaluaion}. These results should be interpreted with a consideration to the nature of the baseline models: the architecture is generic, without physics-informed priors or task-specific tuning. The scores reported---especially those for profile dynamics and MHD activity tasks---highlight benchmark difficulty rather than flaws in optimization. Nevertheless, this baseline establishes a realistic lower bound and identifies areas for improvement.

% ----------------------------------------------------------------------------------
\section{Conclusions}
% ----------------------------------------------------------------------------------

In this work, we introduced \bench{}, the first large-scale, open benchmark specifically designed for evaluating AI models on MAST tokamak diagnostics. We provided a complete open-source training and evaluation stack for 14 diverse downstream tasks together with a collection of four baseline models, creating an integrated framework for benchmarking, tooling, and model development.

\bench\ opens the door to more systematic and reproducible research in fusion plasma modeling. It also provides a platform for exploring advanced representation learning of plasma, short- and long-horizon predictions, and generalization across tokamak operating regimes. We believe the adoption of \bench\ will accelerate progress toward practical, data-driven fusion models, foster stronger collaboration between the fusion and machine learning communities, and ultimately contribute to the development of stable and commercially viable fusion energy.

\small
\bibliographystyle{plainnat}
\bibliography{main}

%%%%%%%%%%%%%%%%%%%%%%%%%%%%%%%%%%%%%%%%%%%%%%%%%%%%%%%%%%%%

\newpage
\appendix

\renewcommand{\thetable}{A.\arabic{table}}
\setcounter{table}{0}

\section{Dataset characterization}

\subsection{Full Set of Signals}

Table~\ref{tab:full_list_of_signals} presents the complete set of signals used in \bench{}. The table is aligned with the taxonomy in Table~\ref{tab:taxonomy} and specifies signals within each category and subcategory. We note that both the data used by \bench\ as well as the FAIR-MAST data are stored in per-shot \texttt{zarr} files. Within each \texttt{zarr} file, signals are organized into groups (which are similar, though not identical, to the categories listed in Table~\ref{tab:full_list_of_signals}). We therefore provide each signal’s full identifier in the format <group name>-<signal name>.

\begin{table}[H]
\centering
\caption{Full set of signals used in \bench.}
\label{tab:full_list_of_signals}
\scriptsize
\renewcommand{\arraystretch}{1.2}
\begin{tabular}{lll}
% \begin{tabular}{p{0.1\linewidth}p{0.1\linewidth}p{0.6\linewidth}}

\toprule
\textbf{Category} & \textbf{Subcategory/Signals} & \textbf{Full signal name} \\
\midrule

\multirow{3}{*}{\textbf{Magnetics}}
& Flux loops   & magnetics-flux\_loop\_flux \\
& Pickup coils & \makecell[l]{magnetics-[b\_field\_pol\_probe\_ccbv\_field, b\_field\_pol\_probe\_obr\_field, \\ b\_field\_pol\_probe\_obv\_field]} \\
& Saddle coils & magnetics-b\_field\_tor\_probe\_saddle\_voltage \\
\midrule

\multirow{2}{*}{\textbf{Kinetics}}
& Thomson scattering & thomson\_scattering-[t\_e, n\_e] \\
& Interferometer     & interferometer-n\_e\_line\\
\midrule

\multirow{2}{*}{\textbf{Radiatives}}
& D$_\text{alpha}$ & spectrometer\_visible-filter\_spectrometer\_dalpha\_voltage\\
& soft X-ray & soft\_x\_rays-[horizontal\_cam\_lower, horizontal\_cam\_upper] \\

\midrule

\textbf{Fast magnetics} &
Mirnov coils &  magnetics-[b\_field\_tor\_probe\_cc\_field, b\_field\_pol\_probe\_omv\_voltage] \\

\midrule

\multirow{3}{*}{\textbf{Currents}}
& Poloidal field coil currents & pf\_active-[coil\_current, solenoid\_current] \\
& Plasma current & summary-ip\\
\midrule

\textbf{Voltages} &
Poloidal field coil voltages & pf\_active-coil\_voltage\\
\midrule

\multirow{2}{*}{\textbf{References}}
& Reference plasma current & pulse\_schedule-i\_plasma\\ 
& Reference plasma density & pulse\_schedule-n\_e\_line\\
\midrule

\multirow{2}{*}{\textbf{Fueling}}
& NBI power   & summary-power\_nbi\\
& Gas puffing & gas\_injection-total\_injected\\
\midrule

\multirow{2}{*}{\textbf{Equilibrium}}
& Shape parameters & \makecell[l]{equilibrium-[elongation, elongation\_axis, triangularity\_upper, triangularity\_lower, \\
x\_point\_r, x\_point\_z, minor\_radius, magnetic\_axis\_r, magnetic\_axis\_z]}\\
& $J_{\text{tor}}$ metrics  & \makecell[l]{equilibrium-[q95, beta\_tor, beta\_pol, beta\_normal, bvac\_rmag, bphi\_rmag]}\\
& plasma boundary  & equilibrium-[lcfs\_r, lcfs\_z]
\\

& flux map         & equilibrium-psi \\
\bottomrule

\end{tabular}
\end{table}

\subsection{Summary Statistics of Diagnostic Signals}
\label{sec:stat_summ}

Table~\ref{tab:compact_summary} provides an overview of the main diagnostic variables used in this work, including their central tendency, variability, and missing-data rates. The table spans equilibrium reconstruction outputs, core plasma parameters, magnetics, fueling and control signals, and diagnostic measurements. This heterogeneous set reflects both physically derived quantities and directly measured signals, with substantially different statistical properties and data quality characteristics across groups. In particular, the reported missingness highlights systematic differences in diagnostic availability, which are further discussed in the following sections. 

Missingness is highly structured and varies significantly by diagnostic modality. 
Equilibrium reconstruction-derived quantities exhibit the highest and most systematic NaN rates, typically around 38–55\%, with missing values concentrated at the beginnings and ends of shots. This behavior is consistent with periods where equilibrium fitting fails to converge or is not well-defined during transient phases of the discharge. 
Thomson scattering measurements also show high missingness but with a different pattern, reflecting intermittent diagnostic coverage and occasional acquisition or processing failures rather than reconstruction instability. 
Other diagnostics, such as magnetics, control system signals, and core operational parameters, generally exhibit low to moderate missingness, although a subset of magnetic probe channels shows elevated dropout rates indicative of partial sensor or channel-level issues.

\begin{table}[H]
\centering
\caption{Compact summary statistics table.}
\label{tab:compact_summary}
\scriptsize

\begin{tabular}{lrrr}
\toprule
\textbf{Full signal name} & \textbf{mean} & \textbf{std} & \textbf{missing (\%)} \\
\midrule

\texttt{equilibrium-beta\_normal} & 1.06 & 0.76 & 38.2 \\
\texttt{equilibrium-beta\_pol} & 0.22 & 0.15 & 38.2 \\
\texttt{equilibrium-beta\_tor} & 3.04 & 2.2 & 38.2 \\
\texttt{equilibrium-bphi\_rmag} & -0.54 & 0.04 & 38.2 \\
\texttt{equilibrium-bvac\_rmag} & -0.47 & 0.05 & 38.2 \\
\texttt{equilibrium-dpressure\_dpsi} & 8.68e+04 & 7.24e+04 & 38.2 \\
\texttt{equilibrium-elongation} & 1.75 & 0.16 & 38.2 \\
\texttt{equilibrium-elongation\_axis} & 1.46 & 0.15 & 38.2 \\
\texttt{equilibrium-f\_df\_dpsi} & 0.4 & 0.35 & 38.2 \\
\texttt{equilibrium-j\_tor} & 7.25e+04 & 1.97e+05 & 38.2 \\
\texttt{equilibrium-lcfs\_r} & 0.74 & 0.38 & 38.2 \\
\texttt{equilibrium-lcfs\_z} & -0.03 & 0.65 & 38.2 \\
\texttt{equilibrium-magnetic\_axis\_r} & 0.85 & 0.08 & 38.2 \\
\texttt{equilibrium-magnetic\_axis\_z} & -0.02 & 0.02 & 38.2 \\
\texttt{equilibrium-minor\_radius} & 0.55 & 0.05 & 38.2 \\
\texttt{equilibrium-psi} & -0.02 & 0.05 & 30.7 \\
\texttt{equilibrium-q95} & 7.72 & 2.55 & 38.2 \\
\texttt{equilibrium-triangularity\_lower} & 0.35 & 0.08 & 38.2 \\
\texttt{equilibrium-triangularity\_upper} & 0.33 & 0.07 & 38.2 \\
\texttt{equilibrium-x\_point\_r} & 0.56 & 0.04 & 55.1 \\
\texttt{equilibrium-x\_point\_z} & -0.05 & 1.1 & 55.2 \\
\texttt{gas\_injection-total\_injected} & 2.51e+21 & 1.79e+21 & 3.5 \\
\texttt{interferometer-n\_e\_line} & 7.97e+19 & 7.73e+19 & 14.9 \\
\texttt{magnetics-b\_field\_pol\_probe\_ccbv\_field} & 0.1 & 0.14 & 4.7 \\
\texttt{magnetics-b\_field\_pol\_probe\_obr\_field} & -0.01 & 0.08 & 6 \\
\texttt{magnetics-b\_field\_pol\_probe\_obv\_field} & -0.03 & 0.03 & 14.4 \\
\texttt{magnetics-b\_field\_pol\_probe\_omv\_voltage} & 6.33 & 160 & 15.3 \\
\texttt{magnetics-b\_field\_tor\_probe\_cc\_field} & 0 & 0 & 27.8 \\
\texttt{magnetics-b\_field\_tor\_probe\_saddle\_voltage} & 0.08 & 0.37 & 23 \\
\texttt{magnetics-flux\_loop\_flux} & -0.27 & 0.38 & 3.5 \\
\texttt{pf\_active-coil\_current} & 708 & 5.04e+03 & 0 \\
\texttt{pf\_active-coil\_voltage} & 14.7 & 557 & 0.2 \\
\texttt{pf\_active-solenoid\_current} & -2.93e+03 & 1.24e+04 & 0 \\
\texttt{pulse\_schedule-i\_plasma} & 4.72e+05 & 2.5e+05 & 0.2 \\
\texttt{pulse\_schedule-n\_e\_line} & 5.39e+19 & 3.07e+19 & 0.2 \\
\texttt{soft\_x\_rays-horizontal\_cam\_lower} & 0 & 0.02 & 17.3 \\
\texttt{soft\_x\_rays-horizontal\_cam\_upper} & 6.1e+26 & 1.96e+27 & 17.3 \\
\texttt{spectrometer\_visible-filter\_spectrometer\_dalpha\_voltage} & 0.36 & 0.39 & 14.9 \\
\texttt{summary-ip} & 3.72e+05 & 2.85e+05 & 0 \\
\texttt{summary-power\_nbi} & 7.03e+05 & 6.61e+05 & 5.4 \\
\texttt{thomson\_scattering-n\_e} & 2.23e+19 & 1.17e+19 & 46.1 \\
\texttt{thomson\_scattering-t\_e} & 424 & 280 & 46.1 \\

\bottomrule
\end{tabular}
\end{table}

% \subsection{Additional dataset information}

% \textbf{Magnetic configurations and operating conditions.} We currently have partial metadata on plasma configurations, with 1,621 annotated shots corresponding to connected double null (CDN) configurations and 568 to lower single null (LSN) configurations.

% \textbf{Plasma-event statistics.} The available metadata includes counts for several key plasma events, including 1,827 H-mode intervals, 819 edge-localised modes (ELMs), 843 internal reconnection events (IREs), 241 locked modes, 152 vertical displacement events (VDEs), 91 back transitions, and 263 sawtooth oscillations.

% \textbf{Rare events and class imbalance.} While events such as VDEs and locked modes are not negligible at the shot level, they become comparatively rare when considered at the level of individual training windows. This results in a strong class imbalance in downstream learning tasks, which increases the difficulty of reliable event detection and classification.

\section{Benchmark characterization}

This section characterizes the benchmark in terms of the number of samples available per task under two evaluation protocols: random splitting and temporal splitting. For each task, we report the number of shots allocated to the training, validation, and test sets. The random split assumes independent and identically distributed samples, whereas the temporal split preserves chronological ordering, resulting in a more challenging and realistic evaluation setting that reflects potential distribution shifts over time.

\renewcommand{\thetable}{B.\arabic{table}}
\setcounter{table}{0}

\begin{table}[H]
\centering
\caption{Task-wise number of shots under Random and Temporal splitting.}
\scriptsize

\begin{tabular}{lccc ccc}
\toprule
& \multicolumn{3}{c}{\textbf{($n_{\text{shots}}$) for random splitting}} & \multicolumn{3}{c}{\textbf{($n_{\text{shots}}$) for temporal splitting}} \\
\cmidrule(lr){2-4} \cmidrule(lr){5-7}
 & Train & Val & Test & Train & Val & Test \\
\midrule

Task 1-1 & 8257 & 1027 & 1020 & 6127 & 1464 & 2712 \\
Task 1-2 & 8382 & 1047 & 1031 & 6237 & 1489 & 2734 \\
Task 1-3 & 8403 & 1050 & 1034 & 6257 & 1492 & 2738 \\
\midrule

Task 2-1 & 8268 & 1028 & 1020 & 6137 & 1465 & 2714 \\
Task 2-2 & 8382 & 1047 & 1031 & 6237 & 1489 & 2734 \\
Task 2-3 & 8403 & 1050 & 1034 & 6257 & 1492 & 2738 \\
\midrule

Task 3-1 & 7226 & 891  & 885  & 5641 & 1357 & 2004 \\
Task 3-2 & 8531 & 1063 & 1060 & 6271 & 1495 & 2889 \\
Task 3-3 & 7072 & 882  & 872  & 5507 & 1331 & 1986 \\
\midrule

Task 4-1 & 8513 & 1065 & 1062 & 6261 & 1494 & 2887 \\
Task 4-2 & 8513 & 1065 & 1062 & 6261 & 1494 & 2887 \\
Task 4-3 & 8062 & 1007 & 992  & 5952 & 1432 & 2687 \\
Task 4-4 & 8912 & 1110 & 1106 & 6473 & 1547 & 3107 \\
Task 4-5 & 5468 & 707  & 671  & 3065 & 765  & 3004 \\
\bottomrule
\end{tabular}

\label{tab:task_random_temporal}
\end{table}

\section{Baselines}
\label{sec:baselines}

\renewcommand{\thetable}{C.\arabic{table}}
\setcounter{table}{0}

\begin{table}[H]
\centering
\caption{\textbf{Model size (number of parameters)} for CNN baseline architectures. Model size is identical across random and temporal splits.}
\scriptsize
\renewcommand{\arraystretch}{0.9}
\setlength{\aboverulesep}{0.5ex}
\setlength{\belowrulesep}{0.5ex}

\begin{tabular}{l cc}
\toprule
\textbf{Task} & \textbf{CNN} & \textbf{CNN+LSTM} \\
\midrule

Task 1-1 & 379,359 & 436,031 \\
Task 1-2 & 274,898 & 331,570 \\
Task 1-3 & 287,377 & 344,049 \\
\midrule

Task 2-1 & 527,178 & 575,658 \\
Task 2-2 & 370,970 & 419,450 \\
Task 2-3 & 383,449 & 431,929 \\
\midrule

Task 3-1 & 304,790 & 291,830 \\
Task 3-2 & 501,693 & 494,877 \\
Task 3-3 & 645,767 & 481,255 \\
\midrule

Task 4-1 & 1,168,206 & 766,126 \\
Task 4-2 & 1,264,466 & 829,618 \\
Task 4-3 & 1,047,053 & 644,973 \\
Task 4-4 & 1,049,165 & 647,085 \\
Task 4-5 & 1,218,702 & 816,622 \\

\bottomrule
\end{tabular}
\label{tab:model_size}
\end{table}

% ----------------------------------------------------------------------------------
\subsection{Multi-branch Convolutional Architecture}
\label{subsec:multi_branch_cnn}
% ----------------------------------------------------------------------------------

\textbf{Preprocessing.} Inputs are standardized using z-score normalization. Remaining missing values are set to zero to obtain fully dense tensors.

\textbf{Window Chunking.} 
Each input and actuator signal window is partitioned into fixed-duration chunks of length 5 ms, using a stride of 5 ms (non-overlapping chunks). Hence, each input variable $v_i \in (T^{(0,v_i)}\times L_h^{(0,v_i)}\times L_w^{(0,v_i)})$ is reshaped into $v_i \in (N_i\times L_t^{(0,v_i)}\times L_h^{(0,v_i)}\times L_w^{(0,v_i)})$ and each actuator $v_{a} \in (T^{(0,v_{a})}\times L_h^{(0,v_{a})}\times L_w^{(0,v_{a})})$ is reshaped into $v_{a} \in (N_{a}\times L_t^{(0,v_{a})}\times L_h^{(0,v_{a})}\times L_w^{(0,v_{a})})$.

\textbf{Encoders.} 
For each input and actuator variable, an independent convolutional encoder is instantiated according to its modality:
1D convolutions for time series, 2D convolutions for profiles, and 3D convolutions for video signals. 
Each encoder consists of a stack of $N$ convolutional layers with kernel size $K$, stride $s$, and padding $p$, followed by ReLU activations, max-pooling, and batch normalization. 
The feature maps are then flattened into a latent vector. The size of the feature map $L_d^{(\ell,v_{i/a})}$ along dimension $d \in (t,h,w)$ after encoder layer $\ell \in [1,N]$ and the flattened latent representation size $L^{(v_{i/a})}$ for variable $v_{i/a}$ can be computed as follows:
% \begin{align}
% L_d^{(\ell,v_i)} &=
%     \left\lfloor
%     \frac{1}{2}
%     \left\lfloor
%         \frac{L_d^{(\ell-1,v_i)} + 2p - K}{s}
%         \right\rfloor + p - \frac{1}{2} 
% \right\rfloor + 1, \\
% L^{(v_i)} &= 2^{N-1}D L_t^{(N,v_i)} L_h^{(N,v_i)} L_w^{(N,v_i)}.
% \end{align}
\begin{align}
L_d^{(\ell,v_{i/a})} &=
    \left\lfloor
    \frac{
    \left\lfloor
        \frac{L_d^{(\ell-1,v_{i/a})} + 2p - K}{s}
        \right\rfloor + 2p - 1 }{2}
\right\rfloor + 1, \\
L^{(v_{i/a})} &= 2^{N-1}D L_t^{(N,v_{i/a})} L_h^{(N,v_{i/a})} L_w^{(N,v_{i/a})}.
\end{align}

\textbf{Decoders.} 

For each output variable $v_o \in (L_t^{(N,v_o)}\times L_h^{(N,v_o)}\times L_w^{(N,v_o)})$, the decoder branch reconstructs the target output from the latent vector, mirroring the encoder structure. 
The shared latent vector is first reshaped into a compressed feature map of dimensionality $L^{(v_o)}$, which is then progressively upsampled using transposed convolutions with the output padding $o_p$. Cropping operations are applied after each transposed convolution to ensure that the reconstructed outputs exactly match the required target dimensions $L_d^{(\ell,v_o)}$. 
% The size feature map size $L_d^{(\ell,v_o)}$ before decoder layer $\ell \in [0, N-1]$ and the compressed flattened feature map are :
The sizes $L_d^{(\ell,v_o)}$ of the feature map before decoder layer $\ell \in [0, N-1]$ and $L^{(v_o)}$ of the compressed flattened feature map are:
\begin{align}
L_d^{(\ell,v_o)}
&=
\left\lceil
s \cdot (L_d^{(\ell+1,v_o)} - 1) - 2p + K + o_p
\right\rceil, \\
L^{(v_o)} &= 2^{N-1}D L_t^{(0,v_o)} L_h^{(0,v_o)} L_w^{(0,v_o)}.
\end{align}

\textbf{Latent Fusion Backbone.}
After encoding each input and actuator variable independently, the resulting latent embeddings are aggregated into a unified representation. Specifically, all modality-specific latent vectors—corresponding to different input variables and actuator signals, are concatenated along the feature dimension to form a joint latent sequence. The concatenated embeddings are then processed by a shared backbone network, which projects the joint representation back to a fixed latent dimensionality $D$, ensuring compatibility with subsequent prediction or decoding modules.

We consider two backbone variants to model interactions across modalities and time:

\textbf{\textit{(i) MLP backbone (the \textbf{CNN Model}).}}
A simple feed-forward architecture composed of stacked linear layers with ReLU activations and dropout. This variant treats the concatenated latent representation as a flattened feature vector and is designed to capture non-linear cross-modal interactions without explicitly modeling temporal order beyond the chunk structure.

\textbf{\textit{(ii) LSTM backbone (the \textbf{CNN+LSTM Model}).}}
A recurrent architecture that processes the concatenated latent embeddings as a sequence of chunk-level tokens. This design explicitly exploits the temporal ordering induced by the chunking procedure, enabling the model to capture temporal dependencies both within and across modalities. The LSTM outputs a refined sequence representation, which is subsequently projected to the target dimensionality $D$.

In both variants, the backbone acts as a fusion and compression stage, transforming heterogeneous latent embeddings into a compact, unified representation suitable for downstream prediction tasks while preserving both cross-modal and temporal structure.

% ----------------------------------------------------------------------------------
\subsection{Experimental Settings}
\label{subsec:exp_settings}
% ----------------------------------------------------------------------------------

\textbf{Parameter Settings.} 
For our multi-branch convolutional architecture, we adopt a latent embedding dimension of $D=16$, with $N=3$ convolutional layers per encoder and decoder blocks.
Each convolution uses a kernel size $K=3$, stride $s=3$, padding $p=1$, and output padding $o_p=1$. 
The same architectural hyperparameters---number of layers, kernel size, stride, and padding---are applied consistently across all tasks.
% , establishing a unified baseline for the benchmark.  

\textbf{Data Preprocessing.}  
We run a set of model-specific preprocessing: first, we standardize data using signal-computed zero mean and unit variance scaler, then we replace NaN values with zeros. This ensures numerical stability and helps to improve training convergence.
The training and validation data is sampled using with a stride of $0.005$ s for Markovian tasks, and $0.025$ s for non-Markovian tasks. Furthermore, because our architecture requires a fixed-length context, inputs for non-Markovian tasks are truncated to a duration of $50$ ms.

\textbf{Training Procedure.} 
We train our models on mini-batches of size 512 using the Adam optimizer with a learning rate of $1\times 10^{-4}$ and early stopping with a patience of 10 steps which terminates training if the validation loss does not improve over this period to prevent overfitting.
We also employ a multi-output mean squared error loss, which averages the loss across all outputs of the model. 
% Table~\ref{tab:task_score} summarizes models sizes per each task.

\textbf{Computational Resources.}
All experiments were run on a single GPU node equipped with an NVIDIA A100-SXM4-80GB GPU (80 GB VRAM) with CUDA 12.8 (driver 570.195.03).

% ==================================================================================
\section{Evaluation}
\label{sec:evaluaion}
% ==================================================================================

\renewcommand{\thetable}{D.\arabic{table}}
\setcounter{table}{0}

In addition to the task- and group-level errors we also report more granular signal-level errors. Moreover, this table includes not only NRMSE-based signal errors but also NMAE-based errors, computed analogously.

\begin{table*}[h]
\centering
\caption{$\textbf{Signal}_{\mathrm{\textbf{NRMSE}}}$ errors for \textbf{random splitting} across all tasks and groups.}
\label{tab:nrmse_random_scores_table}
\small
\renewcommand{\arraystretch}{0.9}

\begin{tabular}{l l l c c c c}

\toprule
& \textbf{Task} & \textbf{Full signal name} & 
$\textbf{Mean}$ & 
$\textbf{Persistence}$ & 
$\textbf{CNN}$ & 
$\textbf{LSTM}$ \\
\midrule

\multirow{18}{*}{\rotatebox[origin=c]{90}{\textbf{Group 1}}} 
& \multirow{15}{*}{1-1}
 & equilibrium-beta\_normal & 0.765 & — & \textbf{0.222} & 0.243 \\
& & equilibrium-beta\_pol & 0.641 & — & \textbf{0.224} & 0.230 \\
& & equilibrium-beta\_tor & 0.844 & — & \textbf{0.225} & 0.267 \\
& & equilibrium-bphi\_rmag & 0.636 & — & \textbf{0.285} & 0.286 \\
& & equilibrium-bvac\_rmag & 0.565 & — & \textbf{0.271} & 0.278 \\
& & equilibrium-elongation & 0.867 & — & \textbf{0.232} & 0.236 \\
& & equilibrium-elongation\_axis & 0.891 & — & \textbf{0.235} & 0.239 \\
& & equilibrium-magnetic\_axis\_r & 0.623 & — & \textbf{0.166} & 0.171 \\
& & equilibrium-magnetic\_axis\_z & 0.746 & — & \textbf{0.122} & 0.152 \\
& & equilibrium-minor\_radius & 0.608 & — & \textbf{0.201} & 0.205 \\
& & equilibrium-q95 & 0.689 & — & \textbf{0.162} & 0.198 \\
& & equilibrium-triangularity\_lower & 0.751 & — & \textbf{0.244} & 0.249 \\
& & equilibrium-triangularity\_upper & 0.889 & — & \textbf{0.239} & 0.243 \\
& & equilibrium-x\_point\_r & 0.815 & — & 0.270 & \textbf{0.255} \\
& & equilibrium-x\_point\_z & 1.018 & — & \textbf{0.039} & 0.070 \\
\cmidrule(lr){2-7}
& \multirow{2}{*}{1-2} 
 & equilibrium-lcfs\_r & 0.996 & — & \textbf{0.054} & 0.058 \\
& & equilibrium-lcfs\_z & 1.001 & — & \textbf{0.040} & 0.044 \\
\cmidrule(lr){2-7}
& 1-3 & equilibrium-psi & 0.927 & — & \textbf{0.148} & 0.159 \\
\midrule

\multirow{21}{*}{\rotatebox[origin=c]{90}{\textbf{Group 2}}}
& \multirow{18}{*}{2-1}
 & equilibrium-beta\_normal & 0.833 & — & 0.359 & \textbf{0.309} \\
& & equilibrium-beta\_pol & 0.717 & — & 0.367 & \textbf{0.318} \\
& & equilibrium-beta\_tor & 0.890 & — & 0.339 & \textbf{0.284} \\
& & equilibrium-bphi\_rmag & 0.716 & — & 0.414 & \textbf{0.357} \\
& & equilibrium-bvac\_rmag & 0.593 & — & 0.324 & \textbf{0.271} \\
& & equilibrium-elongation & 0.874 & — & 0.325 & \textbf{0.254} \\
& & equilibrium-elongation\_axis & 0.917 & — & 0.322 & \textbf{0.254} \\
& & equilibrium-magnetic\_axis\_r & 0.651 & — & 0.202 & \textbf{0.170} \\
& & equilibrium-magnetic\_axis\_z & 0.751 & — & 0.189 & \textbf{0.141} \\
& & equilibrium-minor\_radius & 0.626 & — & 0.253 & \textbf{0.211} \\
& & equilibrium-q95 & 0.727 & — & 0.248 & \textbf{0.184} \\
& & equilibrium-triangularity\_lower & 0.775 & — & 0.305 & \textbf{0.251} \\
& & equilibrium-triangularity\_upper & 0.898 & — & 0.281 & \textbf{0.246} \\
& & equilibrium-x\_point\_r & 0.853 & — & 0.371 & \textbf{0.307} \\
& & equilibrium-x\_point\_z & 1.016 & — & 0.076 & \textbf{0.036} \\
& & pf\_active-coil\_current & 1.042 & — & 0.169 & \textbf{0.133} \\
& & pf\_active-solenoid\_current & 0.854 & — & 0.125 & \textbf{0.088} \\
& & summary-ip & 0.979 & — & 0.282 & \textbf{0.143} \\
\cmidrule(lr){2-7}
& \multirow{2}{*}{2-2}
 & equilibrium-lcfs\_r & 0.992 & — & 0.073 & \textbf{0.064} \\
& & equilibrium-lcfs\_z & 0.991 & — & 0.060 & \textbf{0.049} \\
\cmidrule(lr){2-7}
& 2-3 & equilibrium-psi & 0.923 & — & 0.136 & \textbf{0.126} \\
\midrule

\multirow{10}{*}{\rotatebox[origin=c]{90}{\textbf{Group 3}}}
& \multirow{2}{*}{3-1}
 & thomson\_scattering-n\_e & 0.955 & 0.521 & 0.339 & \textbf{0.314} \\
& & thomson\_scattering-t\_e & 1.012 & 0.552 & 0.453 & \textbf{0.434} \\
\cmidrule(lr){2-7}
& \multirow{3}{*}{3-2}
 & soft\_x\_rays-horizontal\_cam\_lower & 0.581 & \textbf{0.203} & 0.242 & 0.291 \\
& & soft\_x\_rays-horizontal\_cam\_upper & 0.368 & 0.244 & \textbf{0.230} & 0.232 \\
& & spectrometer\_visible-filter\_spectrometer\_dalpha\_voltage & 0.940 & 0.648 & 0.520 & \textbf{0.483} \\
\cmidrule(lr){2-7}
& \multirow{5}{*}{3-3}
 & equilibrium-beta\_normal & 0.801 & — & 0.336 & \textbf{0.289} \\
& & equilibrium-beta\_pol & 0.676 & — & 0.328 & \textbf{0.294} \\
& & equilibrium-beta\_tor & 0.869 & — & 0.303 & \textbf{0.297} \\
& & thomson\_scattering-n\_e & 0.916 & — & 0.557 & \textbf{0.374} \\
& & thomson\_scattering-t\_e & 0.946 & — & 0.406 & \textbf{0.340} \\
\midrule

\multirow{10}{*}{\rotatebox[origin=c]{90}{\textbf{Group 4}}}
& \multirow{2}{*}{4-1}
 & soft\_x\_rays-horizontal\_cam\_lower & 0.623 & 0.320 & 0.266 & \textbf{0.214} \\
& & soft\_x\_rays-horizontal\_cam\_upper & 0.399 & 0.329 & 0.284 & \textbf{0.241} \\
\cmidrule(lr){2-7}
& \multirow{2}{*}{4-2}
 & soft\_x\_rays-horizontal\_cam\_lower & 0.623 & 0.320 & 0.261 & \textbf{0.215} \\
& & soft\_x\_rays-horizontal\_cam\_upper & 0.399 & 0.329 & 0.286 & \textbf{0.242} \\
\cmidrule(lr){2-7}
& 4-3 & equilibrium-magnetic\_axis\_z & 0.731 & — & 0.129 & \textbf{0.118} \\
\cmidrule(lr){2-7}
& 4-4 & summary-ip & 1.027 & 0.718 & 0.530 & \textbf{0.404} \\
\cmidrule(lr){2-7}
& \multirow{2}{*}{4-5} 
 & magnetics-b\_field\_pol\_probe\_omv\_voltage & 0.644 & 0.926 & 0.638 & \textbf{0.632} \\
& & magnetics-b\_field\_tor\_probe\_cc\_field & 0.663 & 0.814 & \textbf{0.663} & 0.665 \\
\bottomrule
\end{tabular}

\end{table*}
\begin{table*}[h]
\centering
\caption{$\textbf{Signal}_{\mathrm{\textbf{NMAE}}}$ errors for \textbf{random splitting} across all tasks and groups.}
\label{tab:nmae_random_scores_table}
\small
\renewcommand{\arraystretch}{0.9}

\begin{tabular}{l l l c c c c}
\toprule
& \textbf{Task} & \textbf{Full signal name} & 
$\textbf{Mean}$ & 
$\textbf{Persistence}$ & 
$\textbf{CNN}$ & 
$\textbf{LSTM}$ \\
\midrule

\multirow{18}{*}{\rotatebox[origin=c]{90}{\textbf{Group 1}}} 
& \multirow{15}{*}{1-1}
 & equilibrium-beta\_normal & 0.648 & — & \textbf{0.159} & 0.182 \\
& & equilibrium-beta\_pol & 0.542 & — & \textbf{0.158} & 0.166 \\
& & equilibrium-beta\_tor & 0.721 & — & \textbf{0.171} & 0.217 \\
& & equilibrium-bphi\_rmag & 0.544 & — & \textbf{0.226} & 0.230 \\
& & equilibrium-bvac\_rmag & 0.511 & — & \textbf{0.239} & 0.248 \\
& & equilibrium-elongation & 0.754 & — & \textbf{0.190} & 0.192 \\
& & equilibrium-elongation\_axis & 0.761 & — & \textbf{0.186} & 0.190 \\
& & equilibrium-magnetic\_axis\_r & 0.561 & — & \textbf{0.134} & 0.139 \\
& & equilibrium-magnetic\_axis\_z & 0.729 & — & \textbf{0.097} & 0.126 \\
& & equilibrium-minor\_radius & 0.550 & — & \textbf{0.162} & 0.164 \\
& & equilibrium-q95 & 0.604 & — & \textbf{0.129} & 0.165 \\
& & equilibrium-triangularity\_lower & 0.682 & — & \textbf{0.195} & 0.196 \\
& & equilibrium-triangularity\_upper & 0.822 & — & 0.194 & \textbf{0.192} \\
& & equilibrium-x\_point\_r & 0.659 & — & 0.189 & \textbf{0.180} \\
& & equilibrium-x\_point\_z & 1.014 & — & \textbf{0.028} & 0.062 \\
\cmidrule(lr){2-7}
& \multirow{2}{*}{1-2} 
 & equilibrium-lcfs\_r & 0.882 & — & \textbf{0.038} & 0.041 \\
& & equilibrium-lcfs\_z & 0.890 & — & \textbf{0.030} & 0.033 \\
\cmidrule(lr){2-7}
& 1-3 & equilibrium-psi & 0.734 & — & \textbf{0.080} & 0.090 \\
\midrule

\multirow{21}{*}{\rotatebox[origin=c]{90}{\textbf{Group 2}}}
& \multirow{18}{*}{2-1}
 & equilibrium-beta\_normal & 0.669 & — & 0.235 & \textbf{0.197} \\
& & equilibrium-beta\_pol & 0.565 & — & 0.239 & \textbf{0.197} \\
& & equilibrium-beta\_tor & 0.738 & — & 0.241 & \textbf{0.195} \\
& & equilibrium-bphi\_rmag & 0.574 & — & 0.298 & \textbf{0.256} \\
& & equilibrium-bvac\_rmag & 0.526 & — & 0.277 & \textbf{0.234} \\
& & equilibrium-elongation & 0.752 & — & 0.253 & \textbf{0.198} \\
& & equilibrium-elongation\_axis & 0.780 & — & 0.249 & \textbf{0.194} \\
& & equilibrium-magnetic\_axis\_r & 0.578 & — & 0.155 & \textbf{0.132} \\
& & equilibrium-magnetic\_axis\_z & 0.727 & — & 0.143 & \textbf{0.104} \\
& & equilibrium-minor\_radius & 0.558 & — & 0.195 & \textbf{0.162} \\
& & equilibrium-q95 & 0.627 & — & 0.195 & \textbf{0.144} \\
& & equilibrium-triangularity\_lower & 0.699 & — & 0.238 & \textbf{0.188} \\
& & equilibrium-triangularity\_upper & 0.827 & — & 0.220 & \textbf{0.194} \\
& & equilibrium-x\_point\_r & 0.681 & — & 0.256 & \textbf{0.207} \\
& & equilibrium-x\_point\_z & 1.011 & — & 0.051 & \textbf{0.024} \\
& & pf\_active-coil\_current & 0.851 & — & 0.120 & \textbf{0.094} \\
& & pf\_active-solenoid\_current & 0.754 & — & 0.101 & \textbf{0.070} \\
& & summary-ip & 0.939 & — & 0.156 & \textbf{0.089} \\
\cmidrule(lr){2-7}
& \multirow{2}{*}{2-2}
 & equilibrium-lcfs\_r & 0.876 & — & 0.049 & \textbf{0.044} \\
& & equilibrium-lcfs\_z & 0.880 & — & 0.044 & \textbf{0.036} \\
\cmidrule(lr){2-7}
& 2-3 & equilibrium-psi & 0.745 & — & 0.078 & \textbf{0.073} \\
\midrule

\multirow{10}{*}{\rotatebox[origin=c]{90}{\textbf{Group 3}}}
& \multirow{2}{*}{3-1}
 & thomson\_scattering-n\_e & 0.806 & 0.358 & 0.247 & \textbf{0.225} \\
& & thomson\_scattering-t\_e & 0.809 & 0.345 & 0.240 & \textbf{0.221} \\
\cmidrule(lr){2-7}
& \multirow{3}{*}{3-2}
 & soft\_x\_rays-horizontal\_cam\_lower & 0.308 & \textbf{0.083} & 0.099 & 0.100 \\
& & soft\_x\_rays-horizontal\_cam\_upper & 0.219 & \textbf{0.101} & 0.114 & 0.119 \\
& & spectrometer\_visible-filter\_spectrometer\_dalpha\_voltage & 0.761 & 0.289 & 0.303 & \textbf{0.270} \\
\cmidrule(lr){2-7}
& \multirow{5}{*}{3-3}
 & equilibrium-beta\_normal & 0.647 & — & 0.229 & \textbf{0.197} \\
& & equilibrium-beta\_pol & 0.536 & — & 0.220 & \textbf{0.193} \\
& & equilibrium-beta\_tor & 0.721 & — & \textbf{0.222} & 0.229 \\
& & thomson\_scattering-n\_e & 0.757 & — & 0.442 & \textbf{0.284} \\
& & thomson\_scattering-t\_e & 0.781 & — & 0.296 & \textbf{0.236} \\
\midrule

\multirow{10}{*}{\rotatebox[origin=c]{90}{\textbf{Group 4}}}
& \multirow{2}{*}{4-1}
 & soft\_x\_rays-horizontal\_cam\_lower & 0.336 & 0.168 & 0.145 & \textbf{0.110} \\
& & soft\_x\_rays-horizontal\_cam\_upper & 0.246 & 0.173 & 0.159 & \textbf{0.129} \\
\cmidrule(lr){2-7}
& \multirow{2}{*}{4-2}
 & soft\_x\_rays-horizontal\_cam\_lower & 0.336 & 0.168 & 0.146 & \textbf{0.113} \\
& & soft\_x\_rays-horizontal\_cam\_upper & 0.246 & 0.173 & 0.159 & \textbf{0.131} \\
\cmidrule(lr){2-7}
& 4-3 & equilibrium-magnetic\_axis\_z & 0.713 & — & 0.096 & \textbf{0.089} \\
\cmidrule(lr){2-7}
& 4-4 & summary-ip & 0.983 & 0.384 & 0.329 & \textbf{0.224} \\
\cmidrule(lr){2-7}
& \multirow{2}{*}{4-5} 
 & magnetics-b\_field\_pol\_probe\_omv\_voltage & \textbf{0.051} & 0.071 & 0.058 & 0.054 \\
& & magnetics-b\_field\_tor\_probe\_cc\_field & 0.011 & \textbf{0.008} & 0.013 & 0.011 \\
\bottomrule
\end{tabular}

\end{table*}

\begin{table*}[h]
\centering
\caption{$\textbf{Signal}_{\mathrm{\textbf{NRMSE}}}$ errors for \textbf{temporal splitting} across all tasks and groups.}
\label{tab:nrmse_temporal_scores_table}
\small
\renewcommand{\arraystretch}{0.9}

\begin{tabular}{l l l c c c c}
\toprule
& \textbf{Task} & \textbf{Full signal name} &
$\textbf{Mean}$ &
$\textbf{Persistence}$ &
$\textbf{CNN}$ &
$\textbf{LSTM}$ \\
\midrule

% ---------------- GROUP 1 ----------------
\multirow{18}{*}{\rotatebox[origin=c]{90}{\textbf{Group 1}}}

& \multirow{15}{*}{1-1}
 & equilibrium-beta\_normal & 1.019 & — & 1.035 & \textbf{0.937} \\
& & equilibrium-beta\_pol & 0.866 & — & 0.906 & \textbf{0.789} \\
& & equilibrium-beta\_tor & 1.075 & — & 0.935 & \textbf{0.913} \\
& & equilibrium-bphi\_rmag & 0.733 & — & 0.728 & \textbf{0.637} \\
& & equilibrium-bvac\_rmag & 0.704 & — & 0.654 & \textbf{0.577} \\
& & equilibrium-elongation & 0.830 & — & \textbf{0.467} & 0.606 \\
& & equilibrium-elongation\_axis & 0.937 & — & \textbf{0.613} & 0.831 \\
& & equilibrium-magnetic\_axis\_r & 0.862 & — & 0.641 & \textbf{0.599} \\
& & equilibrium-magnetic\_axis\_z & 0.915 & — & \textbf{0.409} & 0.415 \\
& & equilibrium-minor\_radius & 0.691 & — & 0.602 & \textbf{0.509} \\
& & equilibrium-q95 & 0.736 & — & 0.422 & \textbf{0.406} \\
& & equilibrium-triangularity\_lower & 0.809 & — & \textbf{0.563} & 0.621 \\
& & equilibrium-triangularity\_upper & 1.080 & — & \textbf{0.540} & 0.602 \\
& & equilibrium-x\_point\_r & 0.916 & — & \textbf{0.702} & 0.809 \\
& & equilibrium-x\_point\_z & 1.016 & — & \textbf{0.068} & 0.097 \\
\cmidrule(lr){2-7}

& \multirow{2}{*}{1-2}
 & equilibrium-lcfs\_r & 1.008 & — & 0.175 & \textbf{0.155} \\
& & equilibrium-lcfs\_z & 0.993 & — & \textbf{0.095} & 0.096 \\
\cmidrule(lr){2-7}

& 1-3 & equilibrium-psi & 0.951 & — & 0.502 & \textbf{0.257} \\
\midrule

% ---------------- GROUP 2 ----------------
\multirow{21}{*}{\rotatebox[origin=c]{90}{\textbf{Group 2}}}

& \multirow{18}{*}{2-1}
 & equilibrium-beta\_normal & 1.064 & — & 2.609 & \textbf{0.921} \\
& & equilibrium-beta\_pol & 0.938 & — & 2.269 & \textbf{0.814} \\
& & equilibrium-beta\_tor & 1.093 & — & 2.125 & \textbf{0.891} \\
& & equilibrium-bphi\_rmag & \textbf{0.771} & — & 1.839 & 0.785 \\
& & equilibrium-bvac\_rmag & 0.718 & — & 1.880 & \textbf{0.581} \\
& & equilibrium-elongation & 0.856 & — & 2.250 & \textbf{0.521} \\
& & equilibrium-elongation\_axis & 0.962 & — & 1.284 & \textbf{0.692} \\
& & equilibrium-magnetic\_axis\_r & 0.870 & — & 1.097 & \textbf{0.629} \\
& & equilibrium-magnetic\_axis\_z & 0.919 & — & 2.141 & \textbf{0.636} \\
& & equilibrium-minor\_radius & 0.702 & — & 2.930 & \textbf{0.585} \\
& & equilibrium-q95 & 0.762 & — & 1.134 & \textbf{0.452} \\
& & equilibrium-triangularity\_lower & 0.829 & — & 3.189 & \textbf{0.457} \\
& & equilibrium-triangularity\_upper & 1.083 & — & 1.394 & \textbf{0.629} \\
& & equilibrium-x\_point\_r & 0.954 & — & 2.147 & \textbf{0.698} \\
& & equilibrium-x\_point\_z & 1.014 & — & 0.404 & \textbf{0.076} \\
& & pf\_active-coil\_current & 0.978 & — & 0.594 & \textbf{0.272} \\
& & pf\_active-solenoid\_current & 0.853 & — & 1.045 & \textbf{0.200} \\
& & summary-ip & 0.773 & — & 1.293 & \textbf{0.235} \\
\cmidrule(lr){2-7}

& \multirow{2}{*}{2-2}
 & equilibrium-lcfs\_r & 1.003 & — & 0.162 & \textbf{0.150} \\
& & equilibrium-lcfs\_z & 0.982 & — & 0.106 & \textbf{0.100} \\
\cmidrule(lr){2-7}

& 2-3 & equilibrium-psi & 0.929 & — & 0.230 & \textbf{0.194} \\
\midrule

% ---------------- GROUP 3 ----------------
\multirow{10}{*}{\rotatebox[origin=c]{90}{\textbf{Group 3}}}

& \multirow{2}{*}{3-1}
 & thomson\_scattering-n\_e & 0.931 & 0.548 & 0.411 & \textbf{0.391} \\
& & thomson\_scattering-t\_e & 1.006 & 0.523 & 0.350 & \textbf{0.325} \\
\cmidrule(lr){2-7}

& \multirow{3}{*}{3-2}
 & soft\_x\_rays-horizontal\_cam\_lower & 1.119 & \textbf{0.104} & 1.899 & 1.940 \\
& & soft\_x\_rays-horizontal\_cam\_upper & 0.240 & \textbf{0.097} & 0.172 & 0.181 \\
& & spectrometer\_visible-filter\_spectrometer\_dalpha\_voltage & 0.951 & \textbf{0.605} & 0.662 & 0.663 \\
\cmidrule(lr){2-7}

& \multirow{5}{*}{3-3}
 & equilibrium-beta\_normal & — & — & \textbf{0.853} & 2.043 \\
& & equilibrium-beta\_pol & — & — & \textbf{0.777} & 2.305 \\
& & equilibrium-beta\_tor & — & — & \textbf{0.786} & 1.487 \\
& & thomson\_scattering-n\_e & — & — & \textbf{0.733} & 0.767 \\
& & thomson\_scattering-t\_e & — & — & \textbf{0.611} & 0.756 \\
\midrule

% ---------------- GROUP 4 ----------------
\multirow{10}{*}{\rotatebox[origin=c]{90}{\textbf{Group 4}}}

& \multirow{2}{*}{4-1}
 & soft\_x\_rays-horizontal\_cam\_lower & 1.132 & \textbf{0.164} & 1.879 & 2.015 \\
& & soft\_x\_rays-horizontal\_cam\_upper & 0.254 & \textbf{0.155} & 0.315 & 0.343 \\
\cmidrule(lr){2-7}

& \multirow{2}{*}{4-2}
 & soft\_x\_rays-horizontal\_cam\_lower & 1.132 & \textbf{0.164} & 4.816 & 2.003 \\
& & soft\_x\_rays-horizontal\_cam\_upper & 0.254 & \textbf{0.155} & 1.304 & 0.286 \\
\cmidrule(lr){2-7}

& 4-3 & equilibrium-magnetic\_axis\_z & 0.863 & — & 1.325 & \textbf{0.820} \\
\cmidrule(lr){2-7}

& 4-4 & summary-ip & 0.924 & 0.639 & 0.579 & \textbf{0.495} \\
\cmidrule(lr){2-7}

& \multirow{2}{*}{4-5}
 & magnetics-b\_field\_pol\_probe\_omv\_voltage & 0.624 & 0.869 & \textbf{0.177} & 0.172 \\
& & magnetics-b\_field\_tor\_probe\_cc\_field & \textbf{1.015} & 1.178 & 1.858 & 1.290 \\
\bottomrule
\end{tabular}

\end{table*}
\begin{table*}[h]
\centering
\caption{$\textbf{Signal}_{\mathrm{\textbf{NMAE}}}$ errors for \textbf{temporal spliting} across all tasks and groups.}
\label{tab:nmae_temporal_scores_table}
\small
\renewcommand{\arraystretch}{0.9}

\begin{tabular}{l l l c c c c}

\toprule
& \textbf{Task} & \textbf{Full signal name} & 
$\textbf{Mean}$ & 
$\textbf{Persistence}$ & 
$\textbf{CNN}$ & 
$\textbf{LSTM}$ \\
\midrule

\multirow{18}{*}{\rotatebox[origin=c]{90}{\textbf{Group 1}}} 
& \multirow{15}{*}{1-1}
& equilibrium-beta\_normal & 0.866 & — & 0.927 & \textbf{0.823} \\
& & equilibrium-beta\_pol & 0.736 & — & 0.801 & \textbf{0.683} \\
& & equilibrium-beta\_tor & 0.914 & — & \textbf{0.837} & 0.799 \\
& & equilibrium-bphi\_rmag & 0.652 & — & 0.634 & \textbf{0.532} \\
& & equilibrium-bvac\_rmag & 0.665 & — & 0.571 & \textbf{0.490} \\
& & equilibrium-elongation & 0.731 & — & \textbf{0.399} & 0.515 \\
& & equilibrium-elongation\_axis & 0.815 & — & \textbf{0.522} & 0.669 \\
& & equilibrium-magnetic\_axis\_r & 0.808 & — & 0.586 & \textbf{0.496} \\
& & equilibrium-magnetic\_axis\_z & 0.890 & — & \textbf{0.340} & 0.357 \\
& & equilibrium-minor\_radius & 0.630 & — & 0.525 & \textbf{0.429} \\
& & equilibrium-q95 & 0.654 & — & 0.356 & \textbf{0.333} \\
& & equilibrium-triangularity\_lower & 0.736 & — & \textbf{0.482} & 0.536 \\
& & equilibrium-triangularity\_upper & 1.011 & — & \textbf{0.470} & 0.523 \\
& & equilibrium-x\_point\_r & 0.740 & — & \textbf{0.584} & 0.646 \\
& & equilibrium-x\_point\_z & 1.012 & — & \textbf{0.051} & 0.076 \\
\cmidrule(lr){2-7}
& \multirow{2}{*}{1-2} 
& equilibrium-lcfs\_r & 0.889 & — & 0.131 & \textbf{0.122} \\
& & equilibrium-lcfs\_z & 0.881 & — & \textbf{0.073} & 0.075 \\
\cmidrule(lr){2-7}
& 1-3 & equilibrium-psi & 0.758 & — & 0.335 & \textbf{0.161} \\
\midrule

\multirow{21}{*}{\rotatebox[origin=c]{90}{\textbf{Group 2}}}
& \multirow{18}{*}{2-1}
& equilibrium-beta\_normal & 0.872 & — & 1.111 & \textbf{0.770} \\
& & equilibrium-beta\_pol & 0.751 & — & 0.951 & \textbf{0.640} \\
& & equilibrium-beta\_tor & 0.913 & — & 0.939 & \textbf{0.765} \\
& & equilibrium-bphi\_rmag & 0.662 & — & 0.760 & \textbf{0.643} \\
& & equilibrium-bvac\_rmag & 0.672 & — & 0.753 & \textbf{0.481} \\
& & equilibrium-elongation & 0.746 & — & 0.940 & \textbf{0.428} \\
& & equilibrium-elongation\_axis & 0.832 & — & 0.824 & \textbf{0.538} \\
& & equilibrium-magnetic\_axis\_r & 0.811 & — & 0.634 & \textbf{0.523} \\
& & equilibrium-magnetic\_axis\_z & 0.888 & — & 0.772 & \textbf{0.533} \\
& & equilibrium-minor\_radius & 0.632 & — & 1.066 & \textbf{0.494} \\
& & equilibrium-q95 & 0.672 & — & 0.544 & \textbf{0.347} \\
& & equilibrium-triangularity\_lower & 0.748 & — & 1.133 & \textbf{0.375} \\
& & equilibrium-triangularity\_upper & 1.012 & — & 0.721 & \textbf{0.525} \\
& & equilibrium-x\_point\_r & 0.761 & — & 0.819 & \textbf{0.539} \\
& & equilibrium-x\_point\_z & 1.008 & — & 0.129 & \textbf{0.048} \\
& & pf\_active-coil\_current & 0.798 & — & 0.231 & \textbf{0.194} \\
& & pf\_active-solenoid\_current & 0.755 & — & 0.317 & \textbf{0.156} \\
& & summary-ip & 0.719 & — & 0.389 & \textbf{0.181} \\
\cmidrule(lr){2-7}
& \multirow{2}{*}{2-2}
& equilibrium-lcfs\_r & 0.884 & — & 0.126 & \textbf{0.114} \\
& & equilibrium-lcfs\_z & 0.870 & — & 0.080 & \textbf{0.078} \\
\cmidrule(lr){2-7}
& 2-3 & equilibrium-psi & 0.759 & — & 0.144 & \textbf{0.127} \\
\midrule

\multirow{10}{*}{\rotatebox[origin=c]{90}{\textbf{Group 3}}}
& \multirow{2}{*}{3-1}
& thomson\_scattering-n\_e & 0.779 & 0.374 & 0.295 & \textbf{0.275} \\
& & thomson\_scattering-t\_e & 0.871 & 0.345 & 0.240 & \textbf{0.222} \\
\cmidrule(lr){2-7}
& \multirow{3}{*}{3-2}
& soft\_x\_rays-horizontal\_cam\_lower & 0.398 & \textbf{0.044} & 0.622 & 0.570 \\
& & soft\_x\_rays-horizontal\_cam\_upper & 0.162 & \textbf{0.041} & 0.097 & 0.104 \\
& & spectrometer\_visible-filter\_spectrometer\_dalpha\_voltage & 0.809 & \textbf{0.252} & 0.413 & 0.443 \\
\cmidrule(lr){2-7}
& \multirow{5}{*}{3-3}
& equilibrium-beta\_normal & 0.861 & — & \textbf{0.662} & 1.711 \\
& & equilibrium-beta\_pol & 0.730 & — & \textbf{0.587} & 1.890 \\
& & equilibrium-beta\_tor & 0.913 & — & \textbf{0.625} & 1.247 \\
& & thomson\_scattering-n\_e & 0.734 & — & \textbf{0.572} & 0.603 \\
& & thomson\_scattering-t\_e & 0.828 & — & \textbf{0.442} & 0.575 \\
\midrule

\multirow{10}{*}{\rotatebox[origin=c]{90}{\textbf{Group 4}}}
& \multirow{2}{*}{4-1}
& soft\_x\_rays-horizontal\_cam\_lower & 0.405 & \textbf{0.088} & 0.743 & 0.959 \\
& & soft\_x\_rays-horizontal\_cam\_upper & 0.171 & \textbf{0.084} & 0.198 & 0.250 \\
\cmidrule(lr){2-7}
& \multirow{2}{*}{4-2}
& soft\_x\_rays-horizontal\_cam\_lower & 0.405 & \textbf{0.088} & 3.271 & 0.918 \\
& & soft\_x\_rays-horizontal\_cam\_upper & 0.171 & \textbf{0.084} & 0.850 & 0.177 \\
\cmidrule(lr){2-7}
& 4-3 & equilibrium-magnetic\_axis\_z & 0.823 & — & 1.254 & \textbf{0.758} \\
\cmidrule(lr){2-7}
& 4-4 & summary-ip & 0.837 & \textbf{0.345} & 0.404 & 0.347 \\
\cmidrule(lr){2-7}
& \multirow{2}{*}{4-5} 
& magnetics-b\_field\_pol\_probe\_omv\_voltage & 0.051 & 0.060 & 0.029 & \textbf{0.022} \\
& & magnetics-b\_field\_tor\_probe\_cc\_field & \textbf{0.012} & 0.013 & 0.078 & 0.025 \\
\bottomrule
\end{tabular}

\end{table*}

% \section{Visualization}

% \newpage
% \input{checklist.tex}

\end{document}